\documentclass[11pt,letterpaper]{article}

\usepackage[a4paper,margin=0.8in,footskip=0.25in]{geometry}
\usepackage{graphicx}
\usepackage{braket}
\usepackage{float}
\usepackage[titletoc,title]{appendix}
\usepackage{multirow}
\usepackage[linesnumbered,ruled,vlined,commentsnumbered]{algorithm2e}
\usepackage{amsfonts}
\usepackage{amsmath}
 \usepackage{mathptmx}

\begin{document}

\begin{center}
{\Large
\textbf\newline\newline\newline{Error tracing in linear and concatenated quantum circuits}
}
\end{center}

\begin{center}
Ritajit Majumdar \textsuperscript{1,*},
Saikat Basu\textsuperscript{2},\\
Priyanka Mukhopadhyay\textsuperscript{3},
Susmita Sur-Kolay\textsuperscript{2}
\\
\bigskip
\textsuperscript{1} Department of Computer Science \& Engineering,\\ University of Calcutta, India\\
* majumdar.ritajit@gmail.com\\
\vspace{0.3cm}
\textsuperscript{2} Advanced Computing \& Microelectronics Unit,\\ Indian Statistical Institute, India\\
\vspace{0.3cm}
\textsuperscript{3} Centre for Quantum Technologies,\\ National University of Singapore\\
\bigskip

\end{center}

\begin{abstract}
Descriptions of quantum algorithms, communication etc. protocols assume the existence of closed quantum system. However, real life quantum systems are open and are highly sensitive to errors. Hence error correction is of utmost importance if quantum computation is to be carried out in reality. Ideally, an error correction block should be placed after every gate operation in a quantum circuit. This increases the overhead and reduced the speedup of the quantum circuit. Moreover, the error correction blocks themselves may induce errors as the gates used for error correction may be noisy. In this paper, we have proposed a procedure to trace error probability due to noisy gates and decoherence in quantum circuit and place an error correcting block only when the error probability exceeds a certain threshold. This procedure shows a drastic reduction in the required number of error correcting blocks. Furthermore, we have considered concatenated codes with tile structure layout lattice architecture\cite{Svore:2007:NTF:2011725.2011727}\cite{spedalieri2008latency},\cite{suchara2013estimating} and SWAP gate based qubit transport mechanism. Tracing errors in higher levels of concatenation shows that, in most cases, after 1 or 2 levels of concatenation, the number of QECC blocks required become static. However, since the gate count increases with increasing concatenation, the percentage saving in gate count is considerably high.
\end{abstract}

\section{Introduction}
\label{sec1}
Quantum computing protocols\cite{PhysRevLett.70.1895,PhysRevLett.69.2881,bennett1984quantum} often assume the presence of an idealised quantum system which is not in contact with its environment and hence is error free. However, it is impossible to attain such idealised situation. So error correction is necessary to allow quantum computation. Quantum error correcting codes\cite{PhysRevA.52.R2493,gottesman1997stabilizer,PhysRevLett.77.793,PhysRevLett.77.198} have been proposed in the literature for this purpose. However, presence of error correcting codes does not prevent the propagation of error in a circuit. Hence a single error, by propagating to different qubits, can cause multiple errors which cannot be corrected by single error correcting codes. Shor\cite{Shor:1996:FQC:874062.875509} has shown that fault tolerant quantum computing is possible. Error correction, together with fault tolerance, ensures reliable quantum computation.\\

However, due to high error affinity of quantum states, repeated error corrections are necessary which often hinders the promised speed of computation and also increases the resource requirement. It may be possible that in small computation, the resource requirement of error correction is more than that of the computation itself. Furthermore, in real situation, the gates used in error correction circuit may be noisy and hence they themselves can incorporate errors in the circuit. Nevertheless, when error probability is low, it is unnecessary to place a QECC block after every gate operation. In this paper, we have proposed a mechanism of error tracing where an offline calculation is done beforehand to trace the error probability through the circuit. When this probability exceeds a predefined threshold, only then an error correcting block is placed. Furthermore, since this process is probabilistic, we have proposed the use fault-tolerant gates to implement error correction. This ensures that even if errors occur, their propagation in the circuit is limited. We have considered this mechanism for both linear as well as concatenated quantum circuits. In both cases the source of error is noisy gates and error due to decoherence, which we call memory error, and Pauli error model has been considered\cite{wilde2013quantum}.\\

The results of our proposed mechanism show a drastic reduction in the required number of QECC blocks, often reducing it to 0. This is, thus, a positive result for performing long quantum computations without having to worry about error correction in every step. If error probability is more, then concatenated codes\cite{nielsen2010quantum} can be used to correct multiple errors. Even though the resource requirement for concatenation increases substantially with increasing level of concatenation, this error tracing mechanism shows that required number of QECC blocks can be reduced by an appreciable amount in concatenated codes too.\\

The remaining paper is organised in the following way - Section 2 gives a brief description of fault tolerance and some of the technologies used to implement a quantum computer. In Section 3, we show the calculation of gate error probability in a linear quantum circuit. Section 4 contains an efficient algorithm to calculate the precise memory error probability in a linear quantum circuit. In Section 5, we combine these two sources of errors to find the total error probability. Section 6 gives a brief review of concatenated codes and we show the application of our proposed technique in concatenated quantum circuits. In Section 7, we provide some benchmark circuits and give the result of the percentage save in QECC blocks by applying our proposed technique. We conclude this paper in Section 8.

\section{Technology and Fault Tolerance}
\label{sec2}
The state of a quantum system is completely described by the wavefunction $\Psi$. Time evolution of this wavefunction is determined by the Schr$\ddot o$dinger equation
\begin{center}
$i\hbar\frac{\partial \ket{\Psi}}{\partial t} = H\ket{\Psi}$
\end{center}

where $\hbar$ is the reduced Plank's Constant = $\frac{h}{2\pi}$ and $H$ is a hermitian operator called the \emph{Hamiltonian} of the system. The most general idea behind quantum computation is to be able to control this Hamiltonian to perform a quantum operation. Different quantum systems have different Hamiltonians and there are also different technologies\cite{6423291} used to implement the computation. An operation may be easily performed in one technology but with difficulty in another. Hence, one technology may be more suitable for implementing a quantum logic gate than another. Different technologies considered in this paper are - Ion Trap (IT)\cite{Roos2014}, Superconductor (SC)\cite{PhysRevLett.91.167005}, Quantum Dot (QD)\cite{PhysRevB.76.035315}, Neutral Atom (NA)\cite{RevModPhys.82.2313}, Linear Photonics (LP)\cite{KnillNature}, Non Linear Photonics (NP)\cite{1464-4266-7-7-002}. The primitive quantum operations realizable in each of the six technologies are shown in table~\ref{tab:pmd}. Since for a particular technology, the non-primitive gates have a higher cost than the primitive ones, it will be useful if after choosing a particular technology, the non-primitive gates of a quantum circuit are realized in terms of the primitive ones.

\begin{table}[h]
\begin{center}
\caption{Supported operations in different technologies\cite{6565421}}
\label{tab:pmd}       
\begin{tabular}{| c | c | c |}
\hline
Technology & One qubit operation & Two qubits operation\\
\hline
QD & $R_x$, $R_z$, X, Z, S, T & CZ\\
NA & $R_{xy}$ & CZ\\
LP & $R_x$, $R_y$, $R_z$, X, Y, Z, S, T, H & CNOT, CZ, SWAP, ZENO\\
NP & Asqu, $R_x$, $R_y$, $R_z$, H & CNOT\\
SC & $R_x$, $R_y$, $R_z$ & iSWAP, CP\\
IT & $R_{xy}$, $R_z$ & G\\
\hline
\end{tabular}
\end{center}
\end{table}

A quantum circuit is said to be fault tolerant if it does not allow error to propagate i.e. if a single error occurs in one of the components, then the procedure will cause at most one error in each of the encoded blocks produced by the component. There exists a group of error correction codes which can be easily implemented fault-tolerantly, called CSS code\cite{PhysRevA.54.1098}. Shor code\cite{PhysRevA.52.R2493} and Steane Code\cite{PhysRevLett.77.793} fall within this group of code. It is shown in \cite{PhysRevLett.77.3260} and \cite{PhysRevA.55.4593} that 5 qubit code\cite{PhysRevLett.77.198} can also be implemented fault-tolerantly.\\

The lattice architectures\cite{Svore:2007:NTF:2011725.2011727},\cite{spedalieri2008latency},\cite{suchara2013estimating} proposed for the error correcting codes support the implementation of gates in Clifford + T library (CTL). However, not every quantum system directly supports the CTL. This makes the implementation of fault-tolerant circuits with CTL inefficient. In \cite{6565421}, the authors have extended the CTL library to a larger set of gates, the fault-tolerant set (FTS), to bridge the gap between CTL and FTS. All the operations in FTS are primitive to each quantum technology and thus can be efficiently implemented. FTS for one qubit is defined as follows
\begin{center}
FTS(1) = \{$R_A(k.\frac{\pi}{4}), H$\} where A $\in$ \{x, y, z\}
\end{center}
Two qubit FTS is defined as follows
\begin{center}
FTS(2) = \{CNOT, CZ, SWAP, ZENO, $G(\frac{\pi}{2})$, $G(\frac{3\pi}{2})$\}
\end{center}
The elements of FTS can be trivially obtained from table ~\ref{tab:fts}.

\begin{table}[h]
\begin{center}
\caption{Conversion between one-qubit FTS and CTL\cite{6565421}}
\label{tab:fts}
\begin{tabular}{| c | c | c | c |}
\hline
$k$ & $R_z$ & $R_x$ & $R_y$\\
\hline
$1$ & $T$ & $HTH$ & $SHTHS^{\dagger}$\\
$2$ & $S$ & $HSH$ & $HZ$\\
$3$ & $ZT^{\dagger}$ & $HZT^{\dagger}H$ & $SHZT^{\dagger}HS^{\dagger}$\\
$4$ & $Z$ & $X$ & $ZX$\\
$5$ & $ZT$ & $HZTH$ & $SHZTHS^{\dagger}$\\
$6$ & $S^{\dagger}$ & $HS^{\dagger}H$ & $ZH$\\
$7$ & $T^{\dagger}$ & $HT^{\dagger}H$ & $SHT^{\dagger}HS^{\dagger}$\\
\hline
\end{tabular}
\end{center}
\end{table}

\section{Tracing error due to noisy gate operation}
\label{sec3}
One of the sources of error in a quantum circuit is noisy gates. In each technology, there are some primitive gates which can be directly implemented (see table ~\ref{tab:pmd}). Other gates are implemented by the combination of these primitive gates. We consider that the gate error probability for the implementation of any primitive gate is $w$. The error probability for non-primitive gates depends on the number of primitive gates required to implement it. We present the fault-tolerant circuits of one gate in each technology considered, as obtained from FTQLS\cite{6565421}, and then calculate the error probability based on these fault-tolerant circuits. The actual value of $w$ may depend on technology and experimental conditions. We have denoted the total gate error probability of the fault tolerant model of each gate by $g_0$. The subscript $0$ indicates that this calculation is for the $0^{th}$ level of concatenation.

\subsection{Ion Trap}
We show the gate error calculation of $R_y$ gate in Ion Trap technology. $R_y$ denotes rotation along y-axis. We have considered a rotation by angle $\pi$. The error probability is given as\\

\begin{figure}[H]
\centering
\includegraphics{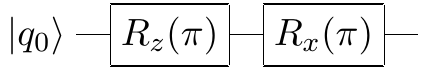}
\label{fig:1}       
\end{figure}

$g_0 = 1 - (1 - g_{0_{Rz}})(1 - g_{0_{Rx}})$\\
\hspace*{5.3ex} $= 1 - (1 - w)(1 - w)$\\
\hspace*{5.3ex} $= 1 - (1 - w)^2$

\subsection{Superconductor}
We consider Geo gate in superconductor technology. The rotation considered is $\frac{\pi}{2}$.

\begin{figure}[H]
\centering
 \includegraphics{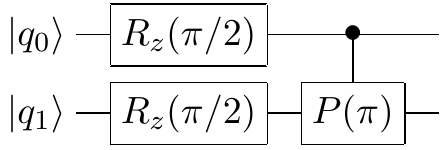}
\label{fig:2}
\end{figure}

$g_0 = 1 - (1 - g_{0_{Rz}})^2(1 - g_{0_{CP}})$\\
\hspace*{5.3ex} $= 1 - (1 - w)^2(1 - w)$\\
\hspace*{5.3ex} $= 1 - (1 - w)^3$

\subsection{Linear Photonics}
In linear photonics, SWAP gate is realised using 3 CNOT gates. The error probability is given as

\begin{figure}[H]
\centering
 \includegraphics{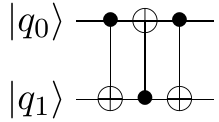}
\label{fig:3}
\end{figure}

$g_0 = 1 - (1 - g_{0_CNOT})^3 = 1 - (1 - w)^3$

\subsection{Non Linear Photonics}
The fault tolerant model and the error probability of Controlled-Z (CZ) gate in Non Linear Photonics is

\begin{figure}[H]
\centering
 \includegraphics{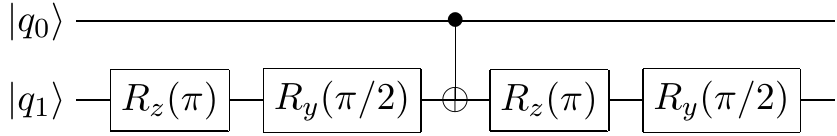}
\label{fig:4}
\end{figure}

$g_0 = 1 - (1 - g_{0_{Rz}})^2(1 - g_{0_{Ry}})^2(1 - g_{0_{CNOT}})$\\
\hspace*{5.3ex} $= 1 - (1 - w)^2(1 - w)^2(1 - w)$\\
\hspace*{5.3ex} $= 1 - (1 - w)^5$

\subsection{Quantum Dot}
The fault tolerant model Zeno gate in Quantum Dot is

\begin{figure}[H]
\centering
 \includegraphics[scale=0.8]{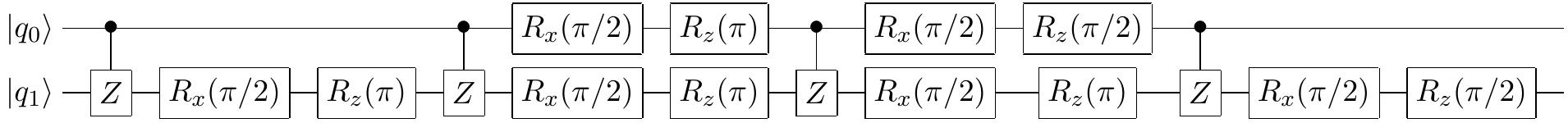}
\label{fig:5}
\end{figure}

$g_0 = 1 - (1 - g_{0_{Rx}})^6(1 - g_{0_{Rz}})^6(1 - g_{0_{CZ}})^4$\\
\hspace*{5.3ex} $= 1 - (1 - w)^6(1 - w)^6(1 - w)^4$\\
\hspace*{5.3ex} $= 1 - (1 - w)^{16}$\\

Table ~\ref{tab:gate_err} contains the error probability of each gate in different technologies.

\begin{table}[h]
\begin{center}
\caption{Comparative study of Gate error probability in different technologies}
\begin{tabular}{| c | c | c | c | c | c | c |}
\hline
Gates & IT & LP & NA & NP & QD & SC\\
\hline
Rx & $w$ & $w$ & $w$ & $w$ & $w$ & $w$\\
Ry & $1 - (1 - w)^2$ & $1 - (1 - w)^2$ & $1 - (1 - w)^2$ & $1 - (1 - w)^2$ & $1 - (1 - w)^3$ & $1 - (1 - w)^2$\\
Rz & $w$ & $w$ & $w$ & $w$ & $w$ & $w$\\
X & $w$ & $w$ & $w$ & $w$ & $w$ & $w$\\
Y & $1 - (1 - w)^2$ & $1 - (1 - w)^2$ & $1 - (1 - w)^2$ & $1 - (1 - w)^2$ & $1 - (1 - w)^3$ & $1 - (1 - w)^2$\\
Z & $w$ & $w$ & $w$ & $w$ & $w$ & $w$\\
H & $1 - (1 - w)^7$ & $1 - (1 - w)^7$ & $1 - (1 - w)^7$ & $1 - (1 - w)^7$ & $1 - (1 - w)^7$ & $1 - (1 - w)^7$\\
S & $w$ & $w$ & $w$ & $w$ & $w$ & $w$\\
T & $w$ & $w$ & $w$ & $w$ & $w$ & $w$\\
CNOT & $1 - (1 - w)^5$ & $w$ & $1 - (1 - w)^3$ & $w$ & $1 - (1 - w)^5$ & $1 - (1 - w)^3$\\
CZ & $1 - (1 - w)^3$ & $w$ & $w$ & $1 - (1 - w)^5$ & $w$ & $w$\\
SWAP & $1 - (1 - w)^{11}$ & $1 - (1 - w)^3$ & $1 - (1 - w)^9$ & $1 - (1 - w)^3$ & $1 - (1 - w)^{16}$ & $1 - (1 - w)^{13}$\\
ZENO & $1 - (1 - w)^{10}$ & $1 - (1 - w)^6$ & $1 - (1 - w)^{12}$ & $1 - (1 - w)^9$ & $1 - (1 - w)^{16}$ & $1 - (1 - w)^{10}$\\
GEO($\frac{\pi}{2}$) & $w$ & $1 - (1 - w)^3$ & $1 - (1 - w)^3$ & $1 - (1 - w)^3$ & $1 - (1 - w)^3$ & $1 - (1 - w)^3$\\
GEO($\frac{3\pi}{4}$) & $1 - (1 - w)^5$ & $1 - (1 - w)^3$ & $1 - (1 - w)^3$ & $1 - (1 - w)^3$ & $1 - (1 - w)^7$ & $1 - (1 - w)^7$\\
GEO($\pi$) & $1 - (1 - w)^5$ & $1 - (1 - w)^2$ & $1 - (1 - w)^5$ & $1 - (1 - w)^2$ & $1 - (1 - w)^9$ & $1 - (1 - w)^2$\\
GEO($\frac{5\pi}{4}$) & $1 - (1 - w)^5$ & $1 - (1 - w)^3$ & $1 - (1 - w)^5$ & $1 - (1 - w)^3$ & $1 - (1 - w)^9$ & $1 - (1 - w)^5$\\
GEO($\frac{3\pi}{2}$) & $1 - (1 - w)^3$ & $1 - (1 - w)^3$ & $1 - (1 - w)^3$ & $1 - (1 - w)^3$ & $1 - (1 - w)^3$ & $1 - (1 - w)^3$\\
\hline
\end{tabular}
\label{tab:gate_err}
\end{center}
\end{table}

\section{Tracing error due to decoherence}
\label{sec4}
Tracing error due to decoherence is not so trivial as noisy gate errors. This is because one requires precise calculation of how long a qubit is idle at any part of the circuit, which on other hand is dependent on the operation times of the quantum gates. Tables ~\ref{tab:gate} and ~\ref{tab:error} contains the data of the time requirement for gate operation and the error probability in each technology considered respectively. We provide an efficient algorithm \ref{mem} which takes as input a .qasm file\cite{6565421}, divides the entire quantum circuit into time slices and calculate which slices are idle. Thus the problem of finding error probability due to decoherence reduces to the problem of finding idle slices. For this algorithm we have assumed that if two gates are operated one after another on a qubit, then the time lapse between them is negligible and decoherence doesn't occur in that time gap. Furthermore, we have considered that no decoherence occurs when a qubit is undergoing a gate operation.

\begin{table}[h]
\begin{center}
\caption{Gate time in ns\cite{suchara2013estimating}}
\label{tab:gate}
\begin{tabular}{| c | c | c | c | c | c | c | c | c | c | c |}
\hline
Technology & CNOT & SWAP & H & $M_x$ & $M_z$ & X & Y & Z & S & T\\
\hline
QD & 27 & 81 & 12 & 100 & 112 & 10 & 11 & 1 & 1 & 1\\
NA & 2533 & 7599 & 781 & 80457 & 80000 & 457 & 457 & 915 & 915 & 915\\
LP & 10 & 10 & 1 & 2 & 1 & 1 & 1 & 1 & 1 & 1\\
NLP & 12 & 36 & 151 & 50 & 1 & 1 & 1 & 1 & 1 & 1\\
SC & 26 & 13 & 16 & 10 & 26 & 10 & 10 & 1 & 1 & 1\\
IT & 120000 & 10000 & 6000 & 106000 & 100000 & 500 & 500 & 3000 & 2000 & 1000\\
\hline
\end{tabular}
\end{center}
\end{table}

\begin{table}[H]
\begin{center}
\caption{Probability of error of worst gate and memory error in different technology\cite{suchara2013estimating}}
\label{tab:error}
\begin{tabular}{| c | c | c |}
\hline
Technology & Probability of Gate error & Memory error (per ns)\\
\hline
QD & $9.89 \times 10^{-1}$ & $3.47 \times 10^{-2}$\\
NA & $8.12 \times 10^{-3}$ & $0.00$\\
LP & $1.01 \times 10^{-1}$ & $9.80 \times 10^{-4}$\\
NLP & $5.20 \times 10^{-3}$ & $9.80 \times 10^{-5}$\\
SC & $1.00 \times 10^{-5}$ & $1.00 \times 10^{-5}$\\
IT & $3.19 \times 10^{-9}$ & $2.52 \times 10^{-12}$\\
\hline
\end{tabular}
\end{center}
\end{table}

\begin{algorithm}[h!]
\SetKwInOut{Input}{input}\SetKwInOut{Output}{output}
\Input{\emph{.qasm} file containing the fault tolerant model of the required circuit as obtained from FTQLS}
\Output{The placement of the gates and the idle regions}
\BlankLine
\Begin{
	\emph{level} $\leftarrow$ list containing the current level of each qubit\;
	\emph{level[i]} $\leftarrow$ 0 for all i\;
	\emph{slice} $\leftarrow$ GCD of the gate operation time of all the gates in the used PMD\;
	\For{each line in \emph{.qasm} file}{
		\If{one qubit gate}{
			\emph{l} $\leftarrow$ level[i], current level of qubit on which the gate operates\;
			Place the gate in level $l+1$\;
			$s \leftarrow$ gate operation time\;
			$sn \leftarrow s/slice$\; \tcc{$sn$ contains the number of slices required} 
			$level[i] = l + sn -1$\; \tcc{-1 since the gates are placed at the current level + 1}
		}
	         \Else{
			$l \leftarrow$ max\{level[i], level[j]\}\;
			Place the gate in level $l+1$ of both the qubits\;
			$s \leftarrow$ gate operation time\;
			$sn \leftarrow s/slice$\;
			level[i] , level[j] = l + sn -1\;
		}
	}
}
\caption{Memory error calculation\label{mem}}
\end{algorithm}

\begin{figure}[h]
\centering
\caption{Example quantum circuit}
\label{fig:circ}
\includegraphics[scale = 1.1]{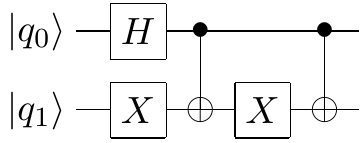}
\end{figure}

FTQLS\cite{6565421} breaks down any multi-qubit gate into one or two qubit gates. Hence we need not worry about gates involving more qubits. We consider a simple circuit shown in figure \ref{fig:circ} and the technology to be quantum dot. We shall use algorithm \ref{mem} to calculate the memory error probability for each qubit in this circuit.\\

The steps for memory error calculation are as follows

\begin{itemize}
\item \emph{slice} = 1 (GCD of time operations for QD gates from Table~\ref{tab:gate})
\item Place H in level 1 (0 + 1) of qubit $\ket{q_0}$ and X in level 1 of qubit $\ket{q_1}$.
\item Gate operation time for H is $s_{H}$ = 12 and that of X is $s_{X}$ = 10.
\item Number of slices required for $\ket{q_0}$ is 12 and that of $\ket{q_1}$ is 10.
\item New \emph{level} of $\ket{q_0}$ is 12 and new \emph{level} of $\ket{q_1}$ is 10.
\item CNOT is a 2 qubit gate. So $l \leftarrow max\{12, 10\}$ = 12.
\item Place CNOT in level $l + 1$ = 13 for both qubits.
\item Qubit $\ket{q_1}$ is idle for 2ns.
\item CNOT acts for 27ns. So new level of both qubits is 13 + 27 - 1 = 39.
\item Place gate X at level 40 for qubit $\ket{q_1}$.
\item X acts for 10ns. So new level of $\ket{q_1}$ is 40 + 10 - 1 = 49.
\item CNOT is a 2 qubit gate. So $l \leftarrow max\{39, 49\}$ = 49.
\item Place CNOT in level 50 for both qubits.
\item Qubit $\ket{q_0}$ is idle for 10ns.
\end{itemize}

Thus after the placement of gates according to the algorithm, we see qubit $\ket{q_0}$ has been idle for 10ns while qubit $\ket{q_1}$ has been idle for 2ns. The memory error probability per ns for QD is $3.47 \times 10^{-2}$. So the error probability in $\ket{q_1}$ is $1 - (1 - 3.47 \times 10^{-2})^2$ = 0.068. Here $(1 - 3.47 \times 10^{-2})$ is the probability of no error in 1ns. Squaring it gives us the probability of no error for 2ns. Finally we subtract it from 1 to have the error probability in 2ns.\\

Similarly, the error probability for qubit $\ket{q_0}$ is $1 - (1 - 3.47 \times 10^{-2})^{10}$ = 0.3. This calculation can be used for simulation before the actual implementation of the quantum circuit.

\section{Error tracing in linear quantum circuit}
\label{sec5}
Calculation of gate error probability for different gates in each technology and tracing memory error have already been taken up individually in the previous sections. Using these two together allows us to trace the total error probability in a linear quantum circuit. We consider again the circuit in figure \ref{fig:circ} to show the steps for error tracing. This calculation is done completely algebraically to make it independent of any technology. The gate error probability for primitive gates is $w$ as before and the probability of \emph{no memory error} for each idle slice is $w_0$.

\begin{itemize}
\item For $\ket{q_0}$, the probability of no error for H is $(1 - w)^7$.
\item For $\ket{q_1}$, the probability of no error for X is $(1 - w)$.
\item From the algorithm to find memory error, we have seen that there are 2 idle slices after X while H is still operating. So the probability of no memory error for two slices is $w_0^2$.
\item Total probability of no error for $\ket{q_1}$ upto this is $w_0^2(1 - w)$.
\item Since CNOT is a 2 qubit gate, we need to update the error probability of both qubits to the maximum of the error probabilities of the qubits. Since $w$ and $w_0$ are both fractions $\ll 1$ and we assume $w_0 \ll w$, we take that $(1 - w)^7 < w_0^2(1 - w)$ (this can change according to actual value of $w$ and $w_0$). If $(1 - w)^7 < w_0^2(1 - w)$, then the error probability of $\ket{q_0}$ is more than that of $\ket{q_1}$. So we update the error probability of both qubits as $1 - (1 - w)^7$.
\item Present probability of no error in both qubits is $(1 - w)^7$.
\item No error probability of CNOT is $(1 - w)^5$. Hence no error probability of both qubits is $(1 - w)^5(1 - w)^7 = (1 - w)^{12}$
\item Qubit $\ket{q_0}$ is now idle for 10ns. So probability of no memory error is $w_0^{10}$.
\item Therefore total probability of no error in $\ket{q_0}$ is $w_0^{10}(1 - w)^{12}$.
\item Probability of no error while X operates on $\ket{q_1}$ is $(1 - w)$.
\item Therefore total probability of no error in $\ket{q_1}$ is $(1 - w)^{13}$.
\item Again considering $w_0 \ll w$, we conclude that $(1 - w)^{13} < w_0^{10}(1 - w)^{12}$.
\item Hence the error probability is more for $\ket{q_1}$.
\item Since CNOT is a two qubit gate,minimum of the no error probability, i.e. $(1 - w)^{13}$ or maximum of the error probability, i.e. $1 - (1 - w)^{13}$ is assigned to both the qubits.
\item Finally, the total error probability of both $\ket{q_0}$ and $\ket{q_1}$ is $1 - (1 - w)^{13}$.
\end{itemize}

Since this calculation is dependent on the actual values of $w$ and $w_0$, the inequality assumptions in steps 5 and 12 may change according to the original values. When the error probability exceeds a pre-defined threshold, we place an error correcting block. The new error probability of the qubits involved is updated to the error probability of the error correcting block.

\section{Error tracing in higher concatenation level}
\label{sec6}
\subsection{Concatenated code}

Though different quantum error correcting codes\cite{PhysRevA.52.R2493,PhysRevLett.77.793,PhysRevLett.77.198} are available in the literature, all these codes assume that the error probability of the quantum channel is low and only a single error can occur. If more than one error occurs, then these codes fail. It is possible to use Stabilizers\cite{gottesman1997stabilizer} to formulate a code which can correct $t$ errors in a system. However, the resource requirement increases largely with the value of $t$ and even that code fails if $t + 1$ errors occur. The solution to this problem is concatenation\cite{nielsen2010quantum}.\\

The information of a single qubit can be protected by distributing it into $n$ qubits. This forms an $n$ qubit error correcting code. We can consider these $n$ physical qubits as a single logical qubit. This logical qubit can again be encoded for error correction using the same or different code. This system is now able to correct 2 errors. By increasing the levels of concatenation in similar way, one can correct any arbitrary number of errors. Further details on concatenated code is available in \cite{nielsen2010quantum}.

\subsection{Tracing of error in higher concatenation}

In sections \ref{sec3}, \ref{sec4} \& \ref{sec5}, we have considered linear circuit, i.e. concatenation level 0. Now, we study the effect of higher concatenation on the error probability. To calculate the error for higher concatenation, an underlying structure for the gate operations is required. We have considered a 2-D nearest neighbour lattice architecture using logical noisy SWAP gates as the basic qubit transport mechanism\cite{Svore:2007:NTF:2011725.2011727,spedalieri2008latency}. In a concatenated architecture, after first level of encoding each physical qubit is replaced by a 2-D block or cell or tile, that forms the logical qubit. After the second level of encoding each such cell is replaced by a 2-D block of such logical qubit cells, and so on. We have considered the tile architecture for the physical layout of Bacon-Shor code\cite{PhysRevA.73.012340,spedalieri2008latency}, Steane code\cite{PhysRevLett.77.793,Svore:2007:NTF:2011725.2011727} and Knill's code\cite{suchara2013estimating}. To the best of our knowledge, the tile structures are possible for the gates in the CTL library or any gate which can be obtained in a compound way from one or more operations in the CTL library.\\

In higher levels of concatenation, when a gate is operated on a logical qubit, some of the qubits in the lower level(s) of concatenation can remain idle even when the gate is operating. So the assumption that during gate operation decoherence cannot occur, does not hold good for higher concatenation level. Let $g_n$ be the gate error at the nth level of encoding, $g_{k_A}$ is the error probability of gate A at $k$-th level of encoding. Similarly let $t_n$ be the gate delay at the $n$-th level of encoding, $t_{k_A}$ is the delay of gate A at $k$-th level of encoding. An error correcting code which can correct a single error fails if two errors occur on the system. For each logical block, probability of failure is
\begin{center}
$Prob$(at least two errors per logical block) = 1 - $Prob$(no gate fails) - $Prob$(exactly one gate fails)
\end{center}
For higher levels of encoding, we consider the failure probability of gates at the next lower level.\\

\begin{figure}[H]
\centering
\caption{An encoded CNOT operation between $\ket{d1,d2,d3,d4}$ (target) and $\ket{q1,q2,q3,q4}$ (control) in Knill code $5 \times 5$ tile architecture}
\label{fig:conc}
\includegraphics[scale = 1]{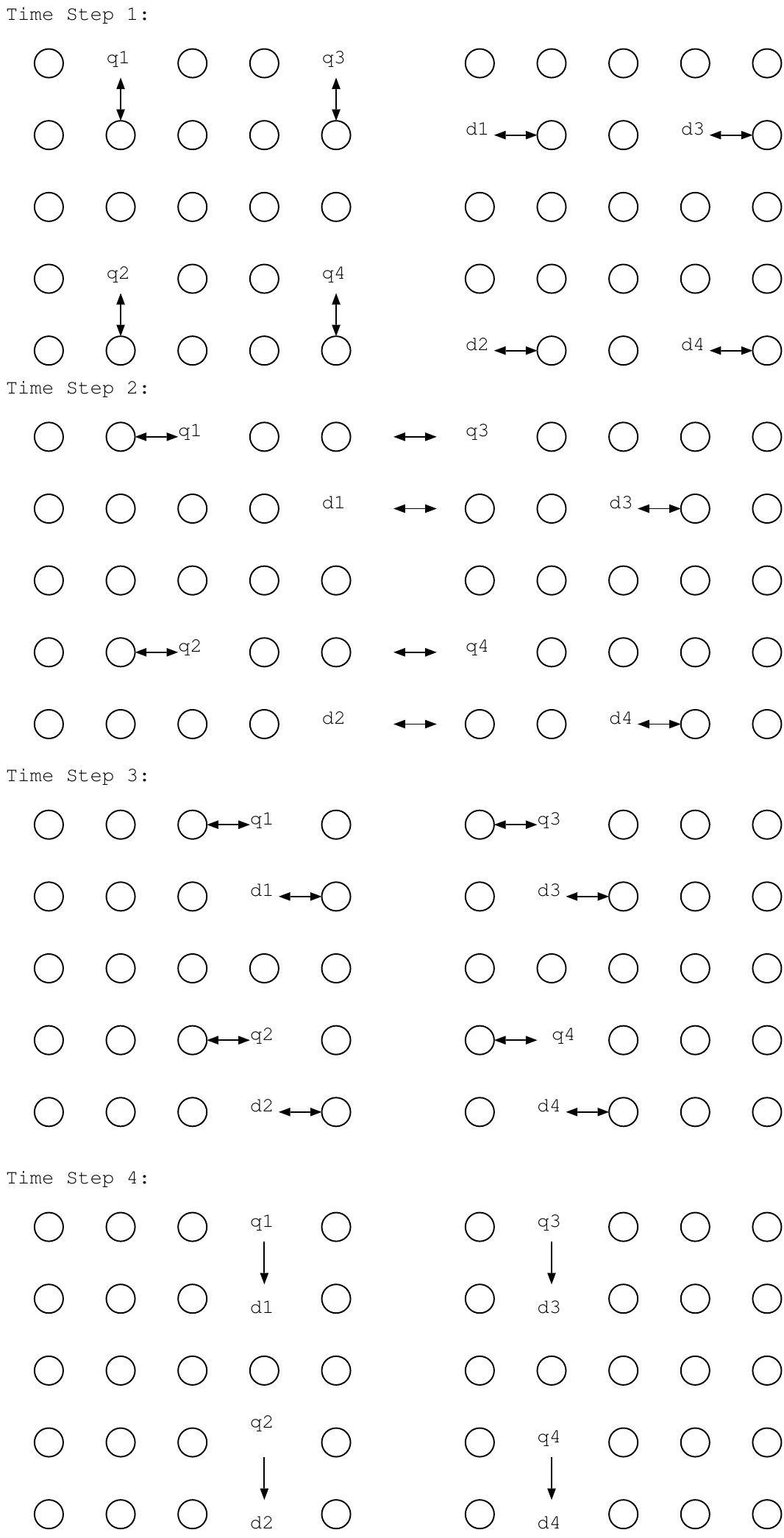}
\end{figure}

Gate error probability in a linear quantum circuit was dependent on the technology used. However, the tile architecture is different for different QECC and hence error probability at the logical level depends on the QECC used and not on the technology. We show the operation of vertical CNOT gate operation in Knill code $5 \times 5$ tile structure in Figure~\ref{fig:conc} and demonstrate the calculation of error. In the figure, $\ket{q}$ represent the set of control qubits and $\ket{d}$ represent the set of target qubits. From \~ref{fig:conc} we can see that the logical CNOT gate at $n^{th}$ level requires 24 SWAP gates and 4 CNOT gates at $(n -1)^{th}$ level. Of these, failure of any one causes errors in at least two qubits. Thus to check the total error probability we need to consider the probability of failure of each gate individually and also both at the same time. Hence the probability of failure at the $n^{th}$ level of encoding is

\begin{multline}
g_n = 1 - (1 -g_{n - 1_{SWAP}} )^{24}(1 - g_{n - 1_{CNOT}} )^4 - \begin{pmatrix}
4\\
1
\end{pmatrix}g_{n - 1_{CNOT}}(1 -g_{n - 1_{SWAP}} )^{24}(1 - g_{n - 1_{CNOT}} )^3\\ - \begin{pmatrix}
24\\
1
\end{pmatrix}g_{n - 1_{SWAP}}(1 -g_{n - 1_{SWAP}} )^{23}(1 - g_{n - 1_{CNOT}} )^4
\end{multline}

In Tables~\ref{tab:log} and ~\ref{tab:log2}, we present the algebraic expressions for the error probability of the logical qubits for various gates in concatenation.

\begin{table}[h]
\caption{Error probability of FTS(2) gates at logical level for each QECC}
\begin{tabular}{| c | p{40mm} | p{40mm} | p{40mm} |}
\hline
Gates & Bacon-Shor & Steane & Knill\\
\hline
SWAP & $1 - (1 - g_{n - 1_{SWAP}})^{33} - \begin{pmatrix}
12\\
1
\end{pmatrix}g_{n - 1_{SWAP}}(1 - g_{n - 1_{SWAP}})^{32}$ & $1 - [(1 - g_{n - 1_{SWAP}})^{26} + 8g_{n - 1_{SWAP}}(1 - g_{n - 1_{SWAP}})^{25}](1 - M_{n_{SWAP}}) - M_{n_{SWAP}} (1 - g_{n - 1_{SWAP}})^{26}$ & $1 - [(1 - g_{n - 1_{SWAP}})^{40} + \begin{pmatrix}
32\\
1
\end{pmatrix}g_{n - 1_{SWAP}}1 - [(1 - g_{n - 1_{SWAP}})^{39}]](1 - M_{n_{SWAP}}) - M_{n_{SWAP}}(1 - g_{n - 1_{SWAP}})^{40}$\\
\hline
CNOT & $1 - (1 - g_{n - 1_{CNOT}})^9(1 - g_{n - 1_{SWAP}})^{54} - \begin{pmatrix}
36\\
1
\end{pmatrix}g_{n - 1_{SWAP}}(1 - g_{n - 1_{CNOT}})^9(1 - g_{n - 1_{SWAP}})^{53} - \begin{pmatrix}
9\\
1
\end{pmatrix}g_{n - 1_{CNOT}}(1 - g_{n - 1_{CNOT}})^8(1 - g_{n - 1_{SWAP}})^{54}$ & $1 - [(1 - g_{n - 1_{SWAP}})^{43}(1 - g_{n - 1_{CNOT}} )^7 + \begin{pmatrix}
26\\
1
\end{pmatrix}g_{n - 1_{SWAP}}(1 - g_{n - 1_{SWAP}})^{42}(1 - g_{n - 1_{CNOT}} )^7 + \begin{pmatrix}
7\\
1
\end{pmatrix}g_{n - 1_{CNOT}}(1 - g_{n - 1_{SWAP}})^{43}(1 - g_{n - 1_{CNOT}} )^6](1 - M_{n_{CNOT}} ) - M_{n_{CNOT}} (1 - g_{n - 1_{SWAP}} )^{43}(1 - g_{n - 1_{CNOT}} )^7$ & $1 - (1 -g_{n - 1_{SWAP}} )^{24}(1 - g_{n - 1_{CNOT}} )^4 - \begin{pmatrix}
4\\
1
\end{pmatrix}g_{n - 1_{CNOT}}(1 -g_{n - 1_{SWAP}} )^{24}(1 - g_{n - 1_{CNOT}} )^3 - \begin{pmatrix}
24\\
1
\end{pmatrix}g_{n - 1_{SWAP}}(1 -g_{n - 1_{SWAP}} )^{23}(1 - g_{n - 1_{CNOT}} )^4$\\
\hline
CZ & $1 - (1 - g_{n_H})(1 - g_{n_{CNOT}} )$ & $1 - (1 - g_{n_H})(1 - g_{n_{CNOT}} )$ & $1 - (1 - g_{n_H})(1 - g_{n_{CNOT}} )$\\
\hline
G & $1 - (1 - g_{n_S})^2(1 - g_{n_H})(1 - g_{n_{CNOT}})$ & $1 - (1 - g_{n_S})^2(1 - g_{n_H})(1 - g_{n_{CNOT}})$ & $1 - (1 - g_{n_S})^2(1 - g_{n_H})(1 - g_{n_{CNOT}})$\\
\hline
ZENO & $1 - (1 - g_{n_S} )^2(1 - g_{n_{SWAP}} )(1 - g_{n_{CZ}} )$ & $1 - (1 - g_{n_S} )^2(1 - g_{n_{SWAP}} )(1 - g_{n_{CZ}} )$ & $1 - (1 - g_{n_S} )^2(1 - g_{n_{SWAP}} )(1 - g_{n_{CZ}} )$\\
\hline
\end{tabular}
\label{tab:log2}
\end{table}

\begin{figure}[H]
\centering
\includegraphics[scale = 1]{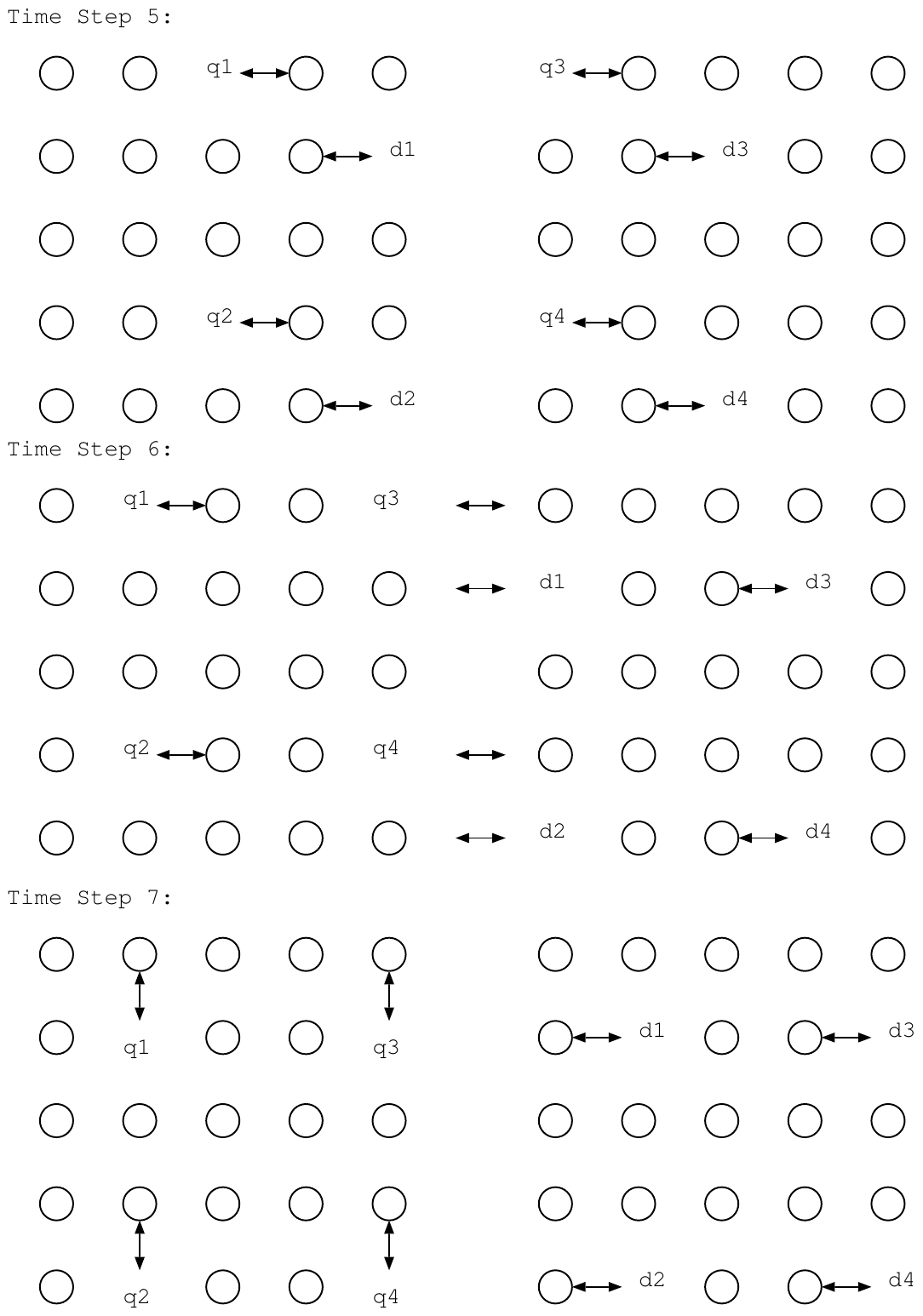}
\end{figure}

\begin{table}[H]
\centering
\footnotesize{
\caption{Error probability of FTS(1) gates at logical level for each QECC}
\label{tab:log}
\begin{tabular}{| c | p{40mm} | p{40mm} | p{45mm} |}
\hline
Gates & Bacon-Shor & Steane & Knill\\
\hline
X & $1 - (1 - g_{n - 1_{X}})^9 -$ \newline $\begin{pmatrix}
9\\
1
\end{pmatrix} g_{n - 1_{X}}(1 - g_{n - 1_{X}})^8$ & $1 - (1 - g_{n - 1_{X}})^7 -$ \newline $\begin{pmatrix}
7\\
1
\end{pmatrix} g_{n - 1_{X}}(1 - g_{n - 1_{X}})^6$ & $1 - (1 - g_{n - 1_{X}})^2 -$ \newline $\begin{pmatrix}
2\\
1
\end{pmatrix} g_{n - 1_{X}}(1 - g_{n - 1_{X}})$\\
\hline
Y & $1 - (1 - g_{n - 1_{Y}})^9 -$ \newline $\begin{pmatrix}
9\\
1
\end{pmatrix} g_{n - 1_{Y}}(1 - g_{n - 1_{Y}})^8$ & $1 - (1 - g_{n - 1_{Y}})^7 -$ \newline $\begin{pmatrix}
7\\
1
\end{pmatrix} g_{n - 1_{Y}}(1 - g_{n - 1_{Y}})^6$ & $1 - (1 - g_{n - 1_{Y}})(1 - g_{n - 1_{X}})(1 - g_{n - 1_{Z}}) - g_{n - 1_{Y}}(1 - g_{n - 1_{X}})(1 - g_{n - 1_{Z}}) - g_{n - 1_{X}}(1 - g_{n - 1_{Y}})(1 - g_{n - 1_{Z}}) - g_{n - 1_{Z}}(1 - g_{n - 1_{X}})(1 - g_{n - 1_{Y}})$\\
\hline
Z & $1 - (1 - g_{n - 1_{Z}})^9 -$ \newline $\begin{pmatrix}
9\\
1
\end{pmatrix} g_{n - 1_{Z}}(1 - g_{n - 1_{Z}})^8$ & $1 - (1 - g_{n - 1_{Z}})^7 -$ \newline $\begin{pmatrix}
7\\
1
\end{pmatrix} g_{n - 1_{Z}}(1 - g_{n - 1_{Z}})^6$ & $1 - (1 - g_{n - 1_{Z}})^2 -$ \newline $\begin{pmatrix}
2\\
1
\end{pmatrix} g_{n - 1_{Z}}(1 - g_{n - 1_{Z}})$\\
\hline
H & $1 - (1 - g_{n - 1_{H}})^9(1 - g_{n - 1_{SWAP}})^{32}(1 - M_{n_H}) - g_{n - 1_{H}}(1 - g_{n - 1_{H}})^8(1 - g_{n - 1_{SWAP}})^{32}(1 - M_{n_{H}}) - M_{n_{H}}(1 - g_{n - 1_{H}})^9(1 - g_{n - 1_{SWAP}})^{32}$ & $1 - (1 - g_{n - 1_{H}})^7 -$ \newline $\begin{pmatrix}
7\\
1
\end{pmatrix} g_{n - 1_{H}}(1 - g_{n - 1_{H}})^6$ & $1 - (1 - g_{n - 1_{H}})^4 -$ \newline $\begin{pmatrix}
4\\
1
\end{pmatrix} g_{n - 1_{H}}(1 - g_{n - 1_{H}})^3$\\
\hline
S & $1 - [(1 - g_{n - 1_{H}})^9(1 - g_{n - 1_{CNOT}})^9(1 - g_{n - 1_{SWAP}})^{42} + \begin{pmatrix}
21\\
1
\end{pmatrix}g_{n - 1_{SWAP}} (1 - g_{n - 1_{SWAP}})^{41}(1 - g_{n - 1_{H}})^9(1 - g_{n - 1_{CNOT}})^9 + \begin{pmatrix}
9\\
1
\end{pmatrix}g_{n - 1_{H}} (1 - g_{n - 1_{SWAP}})^{42}(1 - g_{n - 1_{H}})^8(1 - g_{n - 1_{CNOT}})^9 + \begin{pmatrix}
9\\
1
\end{pmatrix}g_{n - 1_{CNOT}}(1 - g_{n - 1_{SWAP}})^{42}(1 - g_{n - 1_{H}})^9(1 - g_{n - 1_{CNOT}})^8](1 - M_{n_S}) - M_{n_S}(1 - g_{n - 1_{SWAP}})^{42}(1 - g_{n - 1_{H}})^9(1 - g_{n - 1_{CNOT}})^9$ & $1 - (1 - g_{n - 1_{S}})^7(1 - g_{n - 1_{Z}})^7 - \begin{pmatrix}
7\\
1
\end{pmatrix}g_{n - 1_{S}}(1 - g_{n - 1_{S}})^6(1 - g_{n - 1_{Z}})^7$ & $1 - [(1 - g_{n - 1_{SWAP}})^{20}(1 - g_{n - 1_{CNOT}})^4(1 - g_{n - 1_{H}})^4 + \begin{pmatrix}
16\\
1
\end{pmatrix}g_{n - 1_{SWAP}}(1 - g_{n - 1_{SWAP}})^{19}(1 - g_{n - 1_{CNOT}})^4(1 - g_{n - 1_{H}})^4 + \begin{pmatrix}
4\\
1
\end{pmatrix}g_{n - 1_{CNOT}}(1 - g_{n - 1_{SWAP}})^{20}(1 - g_{n - 1_{CNOT}})^{3}(1 - g_{n - 1_{H}})^4 + \begin{pmatrix}
4\\
1
\end{pmatrix}g_{n - 1_{H}}(1 - g_{n - 1_{SWAP}})^{20}(1 - g_{n - 1_{CNOT}})^4(1 - g_{n - 1_{H}})^3](1 - M_{n_{S}}) - M_{n_S}(1 - g_{n - 1_{SWAP}})^{20}(1 - g_{n - 1_{CNOT}})^4(1 - g_{n - 1_{H}})^4$\\
\hline
T & $1 - [(1 - g_{n - 1_{SWAP}})^{42}(1 - g_{n - 1_{CNOT}})^{18}(1 - g_{n - 1_{H}})^9(1 - g_{n - 1_{M_z}})^9 + \begin{pmatrix}
21\\
1
\end{pmatrix}g_{n - 1_{SWAP}}(1 - g_{n - 1_{SWAP}})^{41}(1 - g_{n - 1_{CNOT}})^{18}(1 - g_{n - 1_{H}})^9(1 - g_{n - 1_{M_z}})^9 + \begin{pmatrix}
18\\
1
\end{pmatrix}g_{n - 1_{CNOT}}(1 - g_{n - 1_{SWAP}})^{42}(1 - g_{n - 1_{CNOT}})^{17}(1 - g_{n - 1_{H}})^9(1 - g_{n - 1_{M_z}})^9 + \begin{pmatrix}
9\\
1
\end{pmatrix}g_{n - 1_{H}}(1 - g_{n - 1_{SWAP}})^{42}(1 - g_{n - 1_{CNOT}})^{18}(1 - g_{n - 1_{H}})^7(1 - g_{n - 1_{M_z}})^9 + \begin{pmatrix}
9\\
1
\end{pmatrix}g_{n - 1_{SWAP}}(1 - g_{n - 1_{SWAP}})^{42}(1 - g_{n - 1_{CNOT}})^{18}(1 - g_{n - 1_{H}})^9(1 - g_{n - 1_{M_z}})^8](1 - M_{n_S}) - M_{n_S}(1 - g_{n - 1_{SWAP}})^{42}(1 - g_{n - 1_{CNOT}})^{18}(1 - g_{n - 1_{H}})^9(1 - g_{n - 1_{M_z}})^9$ & $1 - [(1 - g_{n - 1_{SWAP}})^{40}(1 - g_{n - 1_{CNOT}})^7(1 - g_{n - 1_{X}})^7(1- g_{n - 1_{S}})^7(1 - g_{n - 1_{M_z}})^7 + \begin{pmatrix}
22\\
1
\end{pmatrix}g_{n - 1_{SWAP}}(1 - g_{n - 1_{SWAP}})^{39}(1 - g_{n - 1_{CNOT}})^7(1 - g_{n - 1_{X}})^7(1- g_{n - 1_{S}})^7(1 - g_{n - 1_{M_z}})^7 + \begin{pmatrix}
7\\
1
\end{pmatrix}g_{n - 1_{CNOT}}(1 - g_{n - 1_{SWAP}})^{40}(1 - g_{n - 1_{CNOT}})^6(1 - g_{n - 1_{X}})^7(1 - g_{n - 1_{S}})^7(1 - g_{n - 1_{M_z}})^7 + \begin{pmatrix}
7\\
1
\end{pmatrix}g_{n - 1_{M_z}}(1 - g_{n - 1_{SWAP}})^{40}(1 - g_{n - 1_{CNOT}})^7(1 - g_{n - 1_{X}})^7(1 - g_{n - 1_{S}})^7(1 - g_{n - 1_{M_z}})^6 + \begin{pmatrix}
7\\
1
\end{pmatrix}g_{n - 1_{X}}(1 - g_{n - 1_{SWAP}})^{40}(1 - g_{n - 1_{CNOT}})^7(1 - g_{n - 1_{X}})^6(1 - g_{n - 1_{S}})^7(1 - g_{n - 1_{M_z}})^7 + \begin{pmatrix}
7\\
1
\end{pmatrix}g_{n - 1_{S}}(1 - g_{n - 1_{SWAP}})^{40}(1 - g_{n - 1_{CNOT}})^7(1 - g_{n - 1_{X}})^7(1 - g_{n - 1_{S}})^6(1 - g_{n - 1_{M_z}})^7]](1 - M_{n_T}) - M_{n_T}(1 - g_{n - 1_{SWAP}})^{40}(1 -
g_{n - 1_{CNOT} })^7(1 - g_{n - 1_{X}})^7(1- g_{n - 1_{S}})^7(1 - g_{n - 1_{M_z}})^7$ & 1 - [$(1 - g_{n - 1_{SWAP}})^{24}(1 - g_{n - 1_{CNOT}})^8(1 - g_{n - 1_{H}})^4(1 - g_{n - 1_{M_z}})^4 + \begin{pmatrix}
20\\
1
\end{pmatrix} + g_{n - 1_{SWAP}}(1 - g_{n - 1_{SWAP}})^{23}(1 - g_{n - 1_{CNOT} })^8(1 - g_{n - 1_{H}})^4(1- g_{n - 1_{M_z}} )^4 + \begin{pmatrix}
8\\
1
\end{pmatrix}g_{n - 1_{CNOT}}(1- g_{n - 1_{SWAP}})^{24}(1 - g_{n - 1_{CNOT}})^7(1 - g_{n - 1_{H}})^4(1- g_{n - 1_{M_z}})^4 + \begin{pmatrix}
4\\
1
\end{pmatrix}g_{n - 1_{H}}(1 - g_{n - 1_{SWAP}} )^{24}(1 - g_{n - 1_{CNOT}} )^8(1 - g_{n - 1_{H}})^3(1 - g_{n - 1_{M_z}} )^4 + \begin{pmatrix}
4\\
1
\end{pmatrix}g_{n - 1_{M_z}} (1 - g_{n - 1_{SWAP}} )^{24}(1 - g_{n - 1_{CNOT}} )^8(1 - g_{n - 1_{H}})^4(1- g_{n - 1_{M_z}} )^3](1 - M_{n_T} ) - M_{n_T} (1 - g_{n - 1_{SWAP}} )^{24}(1 - g_{n - 1_{CNOT}} )^8(1 - g_{n - 1_{H}})^4(1 - g_{n - 1_{M_z}})4$\\
\hline
\end{tabular}}
\end{table}

\section{Benchmark results}
We have applied our proposed technique of error tracing on various benchmark circuits. For each circuit, we have considered a linear case (concatenation 0) and concatenation level up to 4. Furthermore, for the concatenation codes, we have considered Bacon-Shor, Steane and Knill's code tile architecture for each gate and have varied the threshold as 0.001, 0.01 and 0.1 and have shown the percentage saving in each case.\\

Results of various benchmark circuits are shown in Table~\ref{tab:bvs} - Table~\ref{adder4}. From the values, a few analysis can be made as follows -

\begin{itemize}
\item Ion Trap shows an extremely good result with 100\% savings in lower concatenation levels. Even in higher concatenation levels, the savings are always close to 100\%. However, changing the error threshold does not seem to have much effect in this technology. This is because, the gate error probability in this technology (Table~\ref{tab:error}) is of the order of $10^{-12}$, which is quite lower than the lowest threshold we have considered ($10^{-3}$). For this reason, this technique gives equivalent results in all the three thresholds.
\item Superconductor gives the second best result. Since the error probability in this technology (refer Table~\ref{tab:error}) is more than that of Ion Trap, the savings in this technology is less. Nevertheless, the savings are still near 100\% in most cases. In this technology, both the gate error probability and memory error probability is of the order of $10^{-5}$. So when considering both, the error probability becomes comparable to our thresholds. So here we see that the results are better for higher threshold values.
\item Linear Photonics and Quantum Dot show the worst results when our technique is used. Such results are expected since for LP the gate error probability and memory error probability are of the orders of $10^{-1}$ and $10^{-4}$ respectively and for Quantum Dot they are of the orders of $10^{-1}$ and $10^{-2}$ respectively. We have considered thresholds $10^{-3}, 10^{-2}$ \& $10^{-1}$. Since the error probabilities in these two technologies are always greater than our considered threshold, the results are not very positive in the linear case. However, in higher concatenation levels, there are still much savings.
\item In Non Linear Photonics, the gate error probability is of the order $10^{-3}$ and the memory error probability is of the order $10^{-5}$. Thus, the first threshold, which is $10^{-3}$ cannot provide savings in the linear case as it is same as the gate error probability. However, we see a much better result when the threshold is increased.
\item Neutral atom is similar to Non Linear photonics since its gate error probability is also $10^{-3}$. Similar analysis show that the result will not be very positive in threshold 0.001 but for higher thresholds, savings are observed in the results.
\item For higher concatenation level, memory error plays a big role. This is because even when some gate is operating, qubits in the lower level of concatenation may be idle. This is the reason why with increasing levels of concatenation, the number of required QECC blocks increases initially. However, in higher concatenations, a balance is reached between the idle qubits and the qubits undergoing gate operation in different levels. So, in most cases, the required number of blocks attains an equilibrium. In some tile structures, in higher levels of concatenation, the required number of blocks even decreases. This depends solely on whether some gate is idle or not at any level of concatenation.
\end{itemize}

\begin{table*}[h]
\centering
\caption{Savings on EC blocks for 3 qubit Bernstein Vazirani Search circuit\cite{doi:10.1137/S0097539796300921}}
\label{tab:bvs}
\tiny{
\begin{tabular}{| c | c | c | c | c | c | c | c | c | c |}
\hline
Tech & QECC & Concat & Orig & Th = 0.001 & \% save & Th = 0.01 & \% save & Th = 0.1 & \% save\\
\hline
\multirow{15}{*}{IT} & \multirow{5}{*}{BS} & 0 & 22 & 0 & 100\% & 0 & 100\% & 0 & 100\%\\
& & 1 & 198 & 0 & 100\% & 0 & 100\% & 0 & 100\%\\
& & 2 & 1782 & 15 & 99.16\% & 15 & 99.16\% & 15 & 99.16\%\\
& & 3 & 16038 & 15 & 99.9\% & 15 & 99.9\% & 15 & 99.9\%\\
& & 4 & 144342 & 17 & 99.98\% & 8 & 99.99\% & 0 & 100\%\\
\cline{2-10}
& \multirow{5}{*}{S} & 0 & 22 & 0 & 100\% & 0 & 100\% & 0 & 100\%\\
& & 1 & 154 & 0 & 100\% & 0 & 100\% & 0 & 100\%\\
& & 2 & 1078 & 0 & 100\% & 0 & 100\% & 0 & 100\%\\
& & 3 & 7546 & 0 & 100\% & 0 & 100\% & 0 & 100\%\\
& & 4 & 52822 & 0 & 100\% & 0 & 100\% & 0 & 100\%\\
\cline{2-10}
& \multirow{5}{*}{K} & 0 & 22 & 0 & 100\% & 0 & 100\% & 0 & 100\%\\
& & 1 & 88 & 0 & 100\% & 0 & 100\% & 0 & 100\%\\
& & 2 & 352 & 15 & 95.7\% & 15 & 95.7\% & 15 & 95.7\%\\
& & 3 & 1408 & 15 & 98.9\% & 15 & 98.9\% & 15 & 98.9\%\\
& & 4 & 5632 & 15 & 99.7\% & 15 & 99.7\% & 15 & 99.7\%\\
\hline
\multirow{15}{*}{SC} & \multirow{5}{*}{BS} & 0 & 20 & 0 & 100\% & 0 & 100\% & 0 & 100\%\\
& & 1 & 180 & 3 & 98.3\% & 0 & 100\% & 0 & 100\%\\
& & 2 & 1620 & 17 & 98.9\% & 14 & 99.1\% & 12 & 99.2\%\\
& & 3 & 14580 & 17 & 99.88\% & 17 & 99.88\% & 11 & 99.9\%\\
& & 4 & 131220 & 17 & 99.98\% & 17 & 99.98\% & 17 & 99.98\%\\
\cline{2-10}
& \multirow{5}{*}{S} & 0 & 20 & 0 & 100\% & 0 & 100\% & 0 & 100\%\\
& & 1 & 140 & 0 & 100\% & 0 & 100\% & 0 & 100\%\\
& & 2 & 980 & 2 & 99.79\% & 2 & 99.79\% & 2 & 99.79\%\\
& & 3 & 6860 & 2 & 99.97\% & 2 & 99.97\% & 2 & 99.97\%\\
& & 4 & 48020 & 2 & 99.99\% & 2 & 99.99\% & 2 & 99.99\%\\
\cline{2-10}
& \multirow{5}{*}{K} & 0 & 20 & 0 & 100\% & 0 & 100\% & 0 & 100\%\\
& & 1 & 80 & 0 & 100\% & 0 & 100\% & 0 & 100\%\\
& & 2 & 320 & 14 & 95.6\% & 14 & 95.6\% & 12 & 96.2\%\\
& & 3 & 1280 & 14 & 98.9\% & 14 & 98.9\% & 14 & 98.9\%\\
& & 4 & 5120 & 14 & 99.7\% & 14 & 99.7\% & 14 & 99.7\%\\
\hline
\multirow{15}{*}{LP} & \multirow{5}{*}{BS} & 0 & 22 & 21 & 4.5\% & 21 & 4.5\% & 21 & 4.5\%\\
& & 1 & 198 & 21 & 89.4\% & 21 & 89.4\% & 21 & 89.4\%\\
& & 2 & 1782 & 21 & 98.8\% & 21 & 98.8\% & 21 & 98.8\%\\
& & 3 & 16038 & 21 & 99.8\% & 21 & 99.8\% & 21 & 99.8\%\\
& & 4 & 144342 & 21 & 99.98\% & 21 & 99.98\% & 21 & 99.98\%\\
\cline{2-10}
& \multirow{5}{*}{S} & 0 & 22 & 21 & 4.5\% & 21 & 4.5\% & 21 & 4.5\%\\
& & 1 & 154 & 21 & 86.3\% & 21 & 86.3\% & 21 & 86.3\%\\
& & 2 & 1078 & 21 & 98\% & 21 & 98\% & 21 & 98\%\\
& & 3 & 7546 & 21 & 99.7\% & 21 & 99.7\% & 21 & 99.7\%\\
& & 4 & 52822 & 21 & 99.9\% & 21 & 99.9\% & 21 & 99.9\%\\
\cline{2-10}
& \multirow{5}{*}{K} & 0 & 22 & 21 & 4.5\% & 21 & 4.5\% & 21 & 4.5\%\\
& & 1 & 88 & 21 & 76.1\% & 21 & 76.1\% & 21 & 76.1\%\\
& & 2 & 352 & 21 & 94\% & 21 & 94\% & 21 & 94\%\\
& & 3 & 1408 & 21 & 98.5\% & 21 & 98.5\% & 21 & 98.5\%\\
& & 4 & 5632 & 21 & 99.6\% & 21 & 99.6\% & 21 & 99.6\%\\
\hline
\multirow{15}{*}{NP} & \multirow{5}{*}{BS} & 0 & 22 & 21 & 4.5\% & 9 & 59\% & 0 & 100\%\\
& & 1 & 198 & 19 & 90.4\% & 19 & 90.4\% & 19 & 90.4\%\\
& & 2 & 1782 & 19 & 98.9\% & 19 & 98.9\% & 19 & 98.9\%\\
& & 3 & 16038 & 19 & 99.88\% & 19 & 99.88\% & 19 & 99.88\%\\
& & 4 & 144342 & 19 & 99.98\% & 19 & 99.98\% & 19 & 99.98\%\\
\cline{2-10}
& \multirow{5}{*}{S} & 0 & 22 & 21 & 4.5\% & 9 & 59\% & 0 & 100\%\\
& & 1 & 154 & 19 & 87.66\% & 4 & 97.4\% & 2 & 98.7\%\\
& & 2 & 1078 & 4 & 99.6\% & 2 & 99.8\% & 2 & 99.8\%\\
& & 3 & 7546 & 2 & 99.97\% & 2 & 99.97\% & 2 & 99.97\%\\
& & 4 & 52822 & 2 & 99.99\% & 2 & 99.99\% & 2 & 99.99\%\\
\cline{2-10}
& \multirow{5}{*}{K} & 0 & 22 & 21 & 4.5\% & 9 & 59\% & 0 & 100\%\\
& & 1 & 88 & 14 & 84\% & 14 & 84\% & 6 & 93.1\%\\
& & 2 & 352 & 14 & 96\% & 14 & 96\% & 14 & 96\%\\
& & 3 & 1408 & 14 & 99\% & 14 & 99\% & 14 & 99\%\\
& & 4 & 5632 & 14 & 99.7\% & 14 & 99.7\% & 14 & 99.7\%\\
\hline
\multirow{15}{*}{NA} & \multirow{5}{*}{BS} & 0 & 20 & 19 & 5\% & 8 & 60\% & 1 & 95\%\\
& & 1 & 180 & 19 & 89.4\% & 17 & 90.5\% & 17 & 90.5\%\\
& & 2 & 1620 & 17 & 98.9\% & 17 & 98.9\% & 17 & 98.9\%\\
& & 3 & 14580 & 17 & 99.88\% & 17 & 99.88\% & 17 & 99.88\%\\
& & 4 & 131220 & 17 & 99.99\% & 17 & 99.99\% & 17 & 99.99\%\\
\cline{2-10}
& \multirow{5}{*}{S} & 0 & 20 & 19 & 5\% & 8 & 75\% & 1 & 95\%\\
& & 1 & 140 & 19 & 93.57\% & 11 & 92.1\% & 2 & 98.57\%\\
& & 2 & 980 & 18 & 98.16\% & 3 & 99.69\% & 2 & 99.79\%\\
& & 3 & 6860 & 2 & 99.97\% & 2 & 99.97\% & 2 & 99.97\%\\
& & 4 & 48020 & 2 & 99.99\% & 2 & 99.99\% & 2 & 99.99\%\\
\cline{2-10}
& \multirow{5}{*}{K} & 0 & 20 & 19 & 5\% & 8 & 60\% & 1 & 95\%\\
& & 1 & 80 & 17 & 78.75\% & 14 & 82.5\% & 14 & 82.5\%\\
& & 2 & 320 & 14 & 95.6\% & 14 & 95.6\% & 14 & 95.6\%\\
& & 3 & 1280 & 14 & 98.9\% & 14 & 98.9\% & 14 & 98.9\%\\
& & 4 & 5120 & 14 & 99.7\% & 14 & 99.7\% & 14 & 99.7\%\\
\hline
\multirow{15}{*}{QD} & \multirow{5}{*}{BS} & 0 & 24 & 23 & 4.2\% & 23 & 4.2\% & 23 & 4.2\%\\
& & 1 & 216 & 23 & 89.35\% & 23 & 89.35\% & 23 & 89.35\%\\
& & 2 & 1944 & 23 & 98.8\% & 23 & 98.8\% & 23 & 98.8\%\\
& & 3 & 17496 & 23 & 99.87\% & 23 & 99.87\% & 23 & 99.87\%\\
& & 4 & 157464 & 23 & 99.98\% & 23 & 99.98\% & 23 & 99.98\%\\
\cline{2-10}
& \multirow{5}{*}{S} & 0 & 24 & 23 & 4.2\% & 23 & 4.2\% & 23 & 4.2\%\\
& & 1 & 168 & 23 & 86.3\% & 23 & 86.3\% & 23 & 86.3\%\\
& & 2 & 1176 & 23 & 98\% & 23 & 98\% & 23 & 98\%\\
& & 3 & 8232 & 23 & 99.72\% & 23 & 99.72\% & 23 & 99.72\%\\
& & 4 & 57624 & 23 & 99.96\% & 23 & 99.96\% & 23 & 99.96\%\\
\cline{2-10}
& \multirow{5}{*}{K} & 0 & 24 & 23 & 4.2\% & 23 & 4.2\% & 23 & 4.2\%\\
& & 1 & 96 & 23 & 76\% & 23 & 76\% & 23 & 76\%\\
& & 2 & 384 & 23 & 94\% & 23 & 94\% & 23 & 94\%\\
& & 3 & 1536 & 23 & 98.5\% & 23 & 98.5\% & 23 & 98.5\%\\
& & 4 & 6144 & 23 & 99.62\% & 23 & 99.62\% & 23 & 99.62\%\\
\hline
\end{tabular}
}
\end{table*}

\begin{table}[h]
\centering
\caption{Savings on EC blocks for 4 qubit Quantum Fourier Transform circuit\cite{nielsen2010quantum}}
\label{tab:qft4}
\tiny{
\begin{tabular}{| c | c | c | c | c | c | c | c | c | c |}
\hline
Tech & QECC & Concat & Orig & Th = 0.001 & \% save & Th = 0.01 & \% save & Th = 0.1 & \% save\\
\hline
\multirow{15}{*}{IT} & \multirow{5}{*}{BS} & 0 & 199 & 0 & 100\% & 0 & 100\% & 0 & 100\%\\
& & 1 & 1791 & 0 & 100\% & 0 & 100\% & 0 & 100\%\\
& & 2 & 16119 & 182 & 98.87\% & 182 & 98.87\% & 182 & 98.87\%\\
& & 3 & 145071 & 182 & 99.87\% & 182 & 99.87\% & 182 & 99.87\%\\
& & 4 & 1305639 & 186 & 99.98\% & 186 & 99.98\% & 186 & 99.98\%\\
\cline{2-10}
& \multirow{5}{*}{S} & 0 & 199 & 0 & 100\% & 0 & 100\% & 0 & 100\%\\
& & 1 & 1393 & 0 & 100\% & 0 & 100\% & 0 & 100\%\\
& & 2 & 9751 & 182 & 98.1\% & 182 & 98.1\% & 182 & 98.1\%\\
& & 3 & 68257 & 182 & 99.73\% & 182 & 99.73\% & 182 & 99.73\%\\
& & 4 & 477799 & 182 & 99.96\% & 182 & 99.96\% & 182 & 99.96\%\\
\cline{2 -10}
& \multirow{5}{*}{K} & 0 & 199 & 0 & 100\% & 0 & 100\% & 0 & 100\%\\
& & 1 & 796 & 0 & 100\% & 0 & 100\% & 0 & 100\%\\
& & 2 & 3184 & 182 & 94.28\% & 182 & 94.28\% & 182 & 94.28\%\\
& & 3 & 12736 & 182 & 98.57\% & 182 & 98.57\% & 182 & 98.57\%\\
& & 4 & 50944 & 182 & 99.64\% & 182 & 99.64\% & 182 & 99.64\%\\
\hline
\multirow{15}{*}{SC} & \multirow{5}{*}{BS} & 0 & 261 & 5 & 98.08\% & 0 & 100\% & 0 & 100\%\\
& & 1 & 2349 & 119 & 94.93\% & 15 & 99.36\% & 0 & 100\%\\
& & 2 & 21141 & 260 & 98.77\% & 250 & 98.81\% & 244 & 98.84\%\\
& & 3 & 190269 & 260 & 99.86\% & 260 & 99.86\% & 260 & 99.89\%\\
& & 4 & 1712421 & 260 & 99.98\% & 260 & 99.98\% & 260 & 99.98\%\\
\cline{2-10}
& \multirow{5}{*}{S} & 0 & 261& 5 & 98.08\% & 0 & 100\% & 0 & 100\%\\
& & 1 & 1827 & 52 & 97.15\% & 6 & 99.67\% & 0 & 100\%\\
& & 2 & 12789 & 256 & 97.99\% & 256 & 97.99\% & 256 & 97.99\%\\
& & 3 & 89523 & 256 & 99.71\% & 256 & 99.71\% & 256 & 99.71\%\\
& & 4 & 626661 & 256 & 99.96\% & 256 & 99.96\% & 256 & 99.96\%\\
\cline{2-10}
& \multirow{5}{*}{K} & 0 & 261 & 5 & 98.08\% & 0 & 100\% & 0 & 100\%\\
& & 1 & 1044 & 20 & 98.08\% & 3 & 99.71\% & 0 & 100\%\\
& & 2 & 4176 & 256 & 93.87\% & 256 & 93.87\% & 244 & 94.15\%\\
& & 3 & 16704 & 256 & 98.47\% & 256 & 98.47\% & 256 & 98.47\%\\
& & 4 & 66816 & 256 & 99.62\% & 256 & 99.62\% & 256 & 99.62\%\\
\hline
\multirow{15}{*}{LP} & \multirow{5}{*}{BS} & 0 & 234 & 233 & 0.4\% & 233 & 0.4\% & 233 & 0.4\%\\
& & 1 & 2106 & 233 & 88.93\% & 233 & 88.93\% & 233 & 88.93\%\\
& & 2 & 18954 & 233 & 98.77\% & 233 & 98.77\% & 233 & 98.77\%\\
& & 3 & 170586 & 233 & 99.86\% & 233 & 99.86\% & 233 & 99.86\%\\
& & 4 & 1535274 & 233 & 99.98\% & 233 & 99.98\% & 233 & 99.98\%\\
\cline{2-10}
& \multirow{5}{*}{S} & 0 & 234 & 233 & 0.4\% & 233 & 0.4\% & 233 & 0.4\%\\
& & 1 & 1638 & 233 & 85.77\% & 233 & 85.77\% & 233 & 85.77\%\\
& & 2 & 11466 & 233 & 97.97\% & 233 & 97.97\% & 233 & 97.97\%\\
& & 3 & 80262 & 233 & 99.71\% & 233 & 99.71\% & 233 & 99.71\%\\
& & 4 & 561834 & 233 & 99.96\% & 233 & 99.96\% & 233 & 99.96\%\\
\cline{2-10}
& \multirow{5}{*}{K} & 0 & 234 & 233 & 0.4\% & 233 & 0.4\% & 233 & 0.4\%\\
& & 1 & 936 & 233 & 75.1\% & 233 & 75.1\% & 229 & 75.53\%\\
& & 2 & 3744 & 233 & 93.77\% & 233 & 93.77\% & 229 & 93.88\%\\
& & 3 & 14976 & 233 & 98.44\% & 229 & 98.47\% & 299 & 98.47\%\\
& & 4 & 59904 & 229 & 99.61\% & 229 & 99.61\% & 299 & 99.61\%\\
\hline
\multirow{15}{*}{NP} & \multirow{5}{*}{BS} & 0 & 234 & 233 & 0.4\% & 117 & 50\% & 15 & 93.59\%\\
& & 1 & 2106 & 233 & 88.93\% & 233 & 88.93\% & 233 & 88.93\%\\
& & 2 & 18954 & 233 & 98.77\% & 233 & 98.77\% & 233 & 98.77\%\\
& & 3 & 170586 & 233 & 99.86\% & 233 & 99.86\% & 233 & 99.86\%\\
& & 4 & 1535274 & 233 & 99.98\% & 233 & 99.98\% & 233 & 99.98\%\\
\cline{2-10}
& \multirow{5}{*}{S} & 0 & 234 & 233 & 0.4\% & 117 & 50\% & 15 & 93.59\%\\
& & 1 & 1638 & 233 & 85.77\% & 229 & 86.02\% & 229 & 86.02\%\\
& & 2 & 11466 & 229 & 98\%  & 229 & 98\% & 229 & 98\%\\
& & 3 & 80262 & 229 & 99.71\% & 229 & 99.71\% & 229 & 99.71\%\\
& & 4 & 561834 & 229 & 99.96\% & 229 & 99.96\% & 229 & 99.96\%\\
\cline{2-10}
& \multirow{5}{*}{K} & 0 & 234 & 233 & 0.4\% & 117 & 50\% & 15 & 93.59\%\\
& & 1 & 936 & 229 & 75.53\% & 229 & 75.53\% & 118 & 49.57\%\\
& & 2 & 3744 & 229 & 93.88\% & 229 & 93.88\% & 229 & 93.88\%\\
& & 3 & 14976 & 229 & 98.47\% & 229 & 98.47\% & 229 & 98.47\%\\
& & 4 & 59904 & 229 & 99.62\% & 229 & 99.62\% & 229 & 99.62\%\\
\hline
\multirow{15}{*}{NA} & \multirow{5}{*}{BS} & 0 & 238 & 237 & 0.42\% & 117 & 50.84\% & 18 & 92.44\%\\
& & 1 & 2142 & 237 & 88.93\% & 237 & 88.93\% &237 & 88.93\% \\
& & 2 & 19278 & 237 & 98.77\% & 237 & 98.77\% & 237 & 98.77\%\\
& & 3 & 173502 & 237 & 99.86\% & 237 & 99.86\% & 237 & 99.86\%\\
& & 4 & 1561518 & 237 & 99.98\% & 237 & 99.98\% & 237 & 99.98\%\\
\cline{2-10}
& \multirow{5}{*}{S} & 0 & 238 & 237 & 0.42\% & 117 & 50.84\% & 18 & 92.44\%\\
& & 1 & 1666 & 237 & 85.77\% & 235 & 85.89\% & 228 & 86.31\%\\
& & 2 & 11662 & 237 & 97.97\% & 228 & 98.04\% & 228 & 98.04\%\\
& & 3 & 128282 & 228 & 99.82\% & 228 & 99.82\% & 228 & 99.82\%\\
& & 4 & 897974 & 228 & 99.97\% & 228 & 99.97\% & 228 & 99.97\%\\
\cline{2-10}
& \multirow{5}{*}{K} & 0 & 238 & 237 & 0.42\% & 117 & 50.84\% & 18 & 92.44\%\\
& & 1 & 952 & 237 & 75.1\% & 230 & 75.84\% & 230 & 75.84\%\\
& & 2 & 3808 & 230 & 93.96\% & 230 & 93.96\% & 230 & 93.96\%\\
& & 3 & 15232 & 230 & 98.49\% & 230 & 98.49\% & 230 & 98.49\%\\
& & 4 & 60928 & 230 & 99.62\% & 230 & 99.62\% & 230 & 99.62\%\\
\hline
\multirow{15}{*}{QD} & \multirow{5}{*}{BS} & 0 & 559 & 558 & 0.18\% & 558 & 0.18\% & 558 & 0.18\%\\
& & 1 & 5031 & 558 & 88.91\% & 558 & 88.91\% & 558 & 88.91\%\\
& & 2 & 45279 & 558 & 98.77\% & 558 & 98.77\% & 558 & 98.77\%\\
& & 3 & 407511 & 558 & 99.86\% & 558 & 99.86\% & 558 & 99.86\%\\
& & 4 & 3667599 & 558 & 99.98\% & 558 & 99.98\% & 558 & 99.98\%\\
\cline{2-10}
& \multirow{5}{*}{S} & 0 & 559 & 558 & 0.18\% & 558 & 0.18\% & 558 & 0.18\%\\
& & 1 & 3913 & 558 & 85.74\% & 558 & 85.74\% & 558 & 85.74\%\\
& & 2 & 27391 & 558 & 97.96\% & 558 & 97.96\% & 558 & 97.96\%\\
& & 3 & 191737 & 558 & 99.71\% & 558 & 99.71\% & 558 & 99.71\%\\
& & 4 & 1342159 & 558 & 99.96\% & 558 & 99.96\% & 558 & 99.96\%\\
\cline{2-10}
& \multirow{5}{*}{K} & 0 & 559 & 558 & 0.18\% & 558 & 0.18\% & 558 & 0.18\%\\
& & 1 & 2236 & 558 & 75.04\% & 558 & 75.04\% & 558 & 75.04\%\\
& & 2 & 8944 & 558 & 93.76\% & 558 & 93.76\% & 558 & 93.76\%\\
& & 3 & 35776 & 558 & 98.44\% & 558 & 98.44\% & 558 & 98.44\%\\
& & 4 & 143104 & 558 & 99.61\% & 558 & 99.61\% & 558 & 99.61\%\\
\hline
\end{tabular}
}
\end{table}

\begin{table}[h]
\centering
\caption{Savings on EC blocks for 2 qubit Grover's Search Algorithm circuit\cite{Grover:1996:FQM:237814.237866}}
\label{tab:grover}
\tiny{
\begin{tabular}{| c | c | c | c | c | c | c | c | c | c |}
\hline
Tech & QECC & Concat & Orig & Th = 0.001 & \% save & Th = 0.01 & \% save & Th = 0.1 & \% save\\
\hline
\multirow{15}{*}{IT} & \multirow{5}{*}{BS} & 0 & 22 & 0 & 100\% & 0 & 100\% & 0 & 100\%\\
& & 1 & 198 & 0 & 100\% & 0 & 100\% & 0 & 100\%\\
& & 2 & 1782 & 9 & 99.5\% & 9 & 99.5\% & 9 & 99.5\%\\
& & 3 & 16038 & 9 & 99.94\% & 9 & 99.94\% & 9 & 99.94\%\\
& & 4 & 144342 & 14 & 99.99\% & 9 & 99.99\% & 0 & 100\%\\
\cline{2-10}
& \multirow{5}{*}{S} & 0 & 29 & 0 & 100\% & 0 & 100\% & 0 & 100\%\\
& & 1 & 203 & 0 & 100\% & 0 & 100\% & 0 & 100\%\\
& & 2 & 1421 & 7 & 99.5\% & 7 & 99.5\% & 7 & 99.5\%\\
& & 3 & 9947 & 7 & 99.93\% & 7 & 99.93\% & 7 & 99.93\%\\
& & 4 & 69629 & 7 & 99.99\% & 7 & 99.99\% & 7 & 99.99\%\\
\cline{2-10}
& \multirow{5}{*}{K} & 0 & 29 & 0 & 100\% & 0 & 100\% & 0 & 100\%\\
& & 1 & 116 & 0 & 100\% & 0 & 100\% & 0 & 100\%\\
& & 2 & 464 & 9 & 98.06\% & 9 & 98.06\% & 9 & 98.06\%\\
& & 3 & 1856 & 9 & 99.51\% & 9 & 99.51\% & 9 & 99.51\%\\
& & 4 & 7424 & 9 & 99.88\% & 9 & 99.88\% & 9 & 99.88\%\\
\hline
\multirow{15}{*}{SC} & \multirow{5}{*}{BS} & 0 & 29 & 0 & 100\% & 0 & 100\% & 0 & 100\%\\
& & 1 & 261 & 7 & 97.32\% & 0 & 100\% & 0 & 100\%\\
& & 2 & 2349 & 25 & 98.94\% & 16 & 99.32\% & 9 & 99.62\%\\
& & 3 & 21141 & 25 & 99.88\% & 25 & 99.88\% & 24 & 99.89\%\\
& & 4 & 190269 & 25 & 99.99\% & 25 & 99.99\% & 25 & 99.99\%\\
\cline{2-10}
& \multirow{5}{*}{S} & 0 & 29 & 0 & 100\% & 0 & 100\% & 0 & 100\%\\
& & 1 & 203 & 5 & 97.54\% & 0 & 100\% & 0 & 100\%\\
& & 2 & 1421 & 16 & 98.87\% & 16 & 98.87\% & 16 & 98.87\%\\
& & 3 & 9947 & 16 & 99.84\% & 16 & 99.84\% & 16 & 99.84\%\\
& & 4 & 69629 & 16 & 99.99\% & 16 & 99.99\% & 16 & 99.99\%\\
\cline{2-10}
& \multirow{5}{*}{K} & 0 & 29 & 0 & 100\% & 0 & 100\% & 0 & 100\%\\
& & 1 & 116 & 4 & 96.55\% & 0 & 100\% & 0 & 100\%\\
& & 2 & 464 & 18 & 96.12\% & 18 & 96.12\% & 10 & 97.84\%\\
& & 3 & 1856 & 18 & 99.03\% & 18 & 99.03\% & 18 & 99.03\%\\
& & 4 & 7424 & 18 & 99.78\% & 18 & 99.78\% & 18 & 99.78\%\\
\hline
\multirow{15}{*}{LP} & \multirow{5}{*}{BS} & 0 & 21 & 20 & 4.76\% & 20 & 4.76\% & 20 & 4.76\%\\
& & 1 & 189 & 20 & 89.42\% & 20 & 89.42\% & 20 & 89.42\%\\
& & 2 & 1701 & 20 & 98.82\% & 20 & 98.82\% & 20 & 98.82\%\\
& & 3 & 15309 & 20 & 99.87\% & 20 & 99.87\% & 20 & 99.87\%\\
& & 4 & 137781 & 20 & 99.98\% & 20 & 99.98\% & 20 & 99.98\%\\
\cline{2-10}
& \multirow{5}{*}{S} & 0 & 21 & 20 & 4.76\% & 20 & 4.76\% & 20 & 4.76\%\\
& & 1 & 147 & 20 & 86.39\% & 20 & 86.39\% & 20 & 86.39\%\\
& & 2 & 1029 & 20 & 98.05\% & 20 & 98.05\% & 20 & 98.05\%\\
& & 3 & 7203 & 20 & 99.72\% & 20 & 99.72\% & 20 & 99.72\%\\
& & 4 & 50421 & 20 & 99.96\% & 20 & 99.96\% & 20 & 99.96\%\\
\cline{2-10}
& \multirow{5}{*}{K} & 0 & 21 & 20 & 4.76\% & 20 & 4.76\% & 20 & 4.76\%\\
& & 1 & 84 & 20 & 76.19\% & 20 & 76.19\% & 14 & 83.33\%\\
& & 2 & 336 & 20 & 94.05\% & 20 & 94.05\% & 14 & 95.83\%\\
& & 3 & 1344 & 20 & 98.51\% & 14 & 98.96\% & 14 & 98.96\%\\
& & 4 & 5376 & 14 & 99.74\% & 14 & 99.74\% & 14 & 99.74\%\\
\hline
\multirow{15}{*}{NP} & \multirow{5}{*}{BS} & 0 & 22 & 21 & 4.55\% & 9 & 59.1\% & 0 & 100\%\\
& & 1 & 198 & 18 & 90.9\% & 18 & 90.9\% & 18 & 90.9\%\\
& & 2 & 1782 & 18 & 98.99\% & 18 & 98.99\% & 18 & 98.99\%\\
& & 3 & 16038 & 18 & 99.89\% & 18 & 99.89\% & 18 & 99.89\%\\
& & 4 & 144342 & 18 & 99.99\% & 18 & 99.99\% & 18 & 99.99\%\\
\cline{2-10}
& \multirow{5}{*}{S} & 0 & 22 & 21 & 4.55\% & 9 & 59.1\% & 0 & 100\%\\
& & 1 & 154 & 18 & 88.31\% & 14 & 90.9\% & 14 & 90.9\%\\
& & 2 & 1078 & 14 & 98.7\% & 14 & 98.7\% & 14 & 98.7\%\\
& & 3 & 7546 & 14 & 99.81\% & 14 & 99.81\% & 14 & 99.81\%\\
& & 4 & 52822 & 14 & 99.97\% & 14 & 99.97\% & 14 & 99.97\%\\
\cline{2-10}
& \multirow{5}{*}{K} & 0 & 22 & 21 & 4.55\% & 9 & 59.1\% & 0 & 100\%\\
& & 1 & 88 & 14 & 84.09\% & 14 & 84.09\% & 7 & 92.04\%\\
& & 2 & 352 & 14 & 96.02\% & 14 & 96.02\% & 14 & 96.02\%\\
& & 3 & 1408 & 14 & 99\% & 14 & 99\% & 14 & 99\%\\
& & 4 & 5632 & 14 & 99.75\% & 14 & 99.75\% & 14 & 99.75\%\\
\hline
\multirow{15}{*}{NA} & \multirow{5}{*}{BS} & 0 & 22 & 21 & 4.55\% & 10 & 54.54\% & 1 & 95.45\%\\
& & 1 & 198 & 21 & 89.39\% & 20 & 89.9\% & 20 & 89.9\%\\
& & 2 & 1782 & 20 & 98.88\% & 20 & 98.88\% & 20 & 98.88\%\\
& & 3 & 16038 & 20 & 99.87\% & 20 & 99.87\% & 20 & 99.87\%\\
& & 4 & 144342 & 20 & 99.98\% & 20 & 99.98\% & 20 & 99.98\%\\
\cline{2-10}
& \multirow{5}{*}{S} & 0 & 22 & 21 & 4.55\% & 10 & 54.54\% & 1 & 95.45\%\\
& & 1 & 154 & 21 & 86.36\% & 20 & 87\% & 14 & 90.9\%\\
& & 2 & 1078 & 20 & 98.14\% & 14 & 98.7\% & 14 & 98.7\%\\
& & 3 & 7546 & 14 & 99.81\% & 14 & 99.81\% & 14 & 99.81\%\\
& & 4 & 52822 & 14 & 99.97\% & 14 & 99.97\% & 14 & 99.97\%\\
\cline{2-10}
& \multirow{5}{*}{K} & 0 & 22 & 21 & 4.55\% & 10 & 54.54\% & 1 & 95.45\%\\
& & 1 & 88 & 20 & 77.27\% & 14 & 84.09\% & 14 & 84.09\%\\
& & 2 & 352 & 14 & 96.02\% & 14 & 96.02\% & 14 & 96.02\%\\
& & 3 & 1408 & 14 & 99\% & 14 & 99\% & 14 & 99\%\\
& & 4 & 5632 & 14 & 99.75\% & 14 & 99.75\% & 14 & 99.75\%\\
\hline
\multirow{15}{*}{QD} & \multirow{5}{*}{BS} & 0 & 34 & 33 & 2.94\% & 33 & 2.94\% & 33 & 2.94\%\\
& & 1 & 306 & 33 & 89.21\% & 33 & 89.21\% & 33 & 89.21\%\\
& & 2 & 2754 & 33 & 98.8\% & 33 & 98.8\% & 33 & 98.8\%\\
& & 3 & 24786 & 33 & 99.87\% & 33 & 99.87\% & 33 & 99.87\%\\
& & 4 & 223074 & 33 & 99.98\% & 33 & 99.98\% & 33 & 99.98\%\\
\cline{2-10}
& \multirow{5}{*}{S} & 0 & 34 & 33 & 2.94\% & 33 & 2.94\% & 33 & 2.94\%\\
& & 1 & 238 & 33 & 86.13\% & 33 & 86.13\% & 33 & 86.13\%\\
& & 2 & 1666 & 33 & 98\% & 33 & 98\% & 33 & 98\%\\
& & 3 & 11662 & 33 & 99.72\% & 33 & 99.72\% & 33 & 99.72\%\\
& & 4 & 81634 & 33 & 99.96\% & 33 & 99.96\% & 33 & 99.96\%\\
\cline{2-10}
& \multirow{5}{*}{K} & 0 & 34 & 33 & 2.94\% & 33 & 2.94\% & 33 & 2.94\%\\
& & 1 & 136 & 33 & 75.73\% & 33 & 75.73\% & 33 & 75.73\%\\
& & 2 & 544 & 33 & 93.93\% & 33 & 93.93\% & 33 & 93.93\%\\
& & 3 & 2176 & 33 & 98.48\% & 33 & 98.48\% & 33 & 98.48\%\\
& & 4 & 8704 & 33 & 99.62\% & 33 & 99.62\% & 33 & 99.62\%\\
\hline
\end{tabular}
}
\end{table}

\begin{table}[h]
\centering
\caption{Savings on EC blocks for 4 qubit Reversible Adder circuit}
\label{tab:adder4}
\tiny{
\begin{tabular}{| c | c | c | c | c | c | c | c | c | c |}
\hline
Tech & QECC & Concat & Orig & Th = 0.001 & \% save & Th = 0.01 & \% save & Th = 0.1 & \% save\\
\hline
\multirow{15}{*}{IT} & \multirow{5}{*}{BS} & 0 & 134 & 0 & 100\% & 0 & 100\% & 0 & 100\%\\
& & 1 & 1206 & 0 & 100\% & 0 & 100\% & 0 & 100\%\\
& & 2 & 10854 & 72 & 99.34\% & 72 & 99.34\% & 72 & 99.34\%\\
& & 3 & 97686 & 72 & 99.92\% & 72 & 99.92\% & 72 & 99.92\%\\
& & 4 & 879174 & 76 & 99.99\% & 66 & 99.99\% & 4 & 99.999\%\\
\cline{2-10}
& \multirow{5}{*}{S} & 0 & 134 & 0 & 100\% & 0 & 100\% & 0 & 100\%\\
& & 1 & 938 & 0 & 100\% & 0 & 100\% & 0 & 100\%\\
& & 2 & 6566 & 35 & 99.47\% & 35 & 99.47\% & 35 & 99.47\%\\
& & 3 & 45962 & 35 & 99.92\% & 35 & 99.92\% & 35 & 99.92\%\\
& & 4 & 321734 & 35 & 99.98\% & 35 & 99.98\% & 35 & 99.98\%\\
\cline{2-10}
& \multirow{5}{*}{K} & 0 & 134 & 0 & 100\% & 0 & 100\% & 0 & 100\%\\
& & 1 & 536 & 0 & 100\% & 0 & 100\% & 0 & 100\%\\
& & 2 & 2144 & 72 & 96.64\% & 72 & 96.64\% & 72 & 96.64\%\\
& & 3 & 8576 & 72 & 99.16\% & 72 & 99.16\% & 72 & 99.16\%\\
& & 4 & 34304 & 72 & 99.79\% & 72 & 99.79\% & 72 & 99.79\%\\
\hline
\multirow{15}{*}{SC} & \multirow{5}{*}{BS} & 0 & 183 & 2 & 98.91\% & 0 & 100\% & 0 & 100\%\\
& & 1 & 1647 & 58 & 96.48\% & 9 & 99.45\% & 0 & 100\%\\
& & 2 & 14823 & 168 & 98.87\% & 110 & 99.26\% & 62 & 99.58\%\\
& & 3 & 133407 & 168 & 99.87\% & 168 & 99.87\% & 165 & 99.88\%\\
& & 4 & 1200663 & 168 & 99.98\% & 168 & 99.98\% & 168 & 99.98\%\\
\cline{2-10}
& \multirow{5}{*}{S} & 0 & 183 & 2 & 98.91\% & 0 & 100\% & 0 & 100\%\\
& & 1 & 1281 & 34 & 97.34\% & 4 & 99.69\% & 0 & 100\%\\
& & 2 & 8967 & 118 & 98.68\% & 118 & 98.68\% & 118 & 98.68\%\\
& & 3 & 62769 & 118 & 99.81\% & 118 & 99.81\% & 118 & 99.81\%\\
& & 4 & 439383 & 118 & 99.97\% & 118 & 99.97\% & 118 & 99.97\%\\
\cline{2-10}
& \multirow{5}{*}{K} & 0 & 183 & 2 & 98.91\% & 0 & 100\% & 0 & 100\%\\
& & 1 & 732 & 32 & 95.63\% & 7 & 99\% & 0 & 100\%\\
& & 2 & 2928 & 129 & 95.6\% & 129 & 95.6\% & 65 & 97.78\%\\
& & 3 & 11712 & 129 & 98.9\% & 129 & 98.9\% & 129 & 98.9\%\\
& & 4 & 46848 & 129 & 99.72\% & 129 & 99.72\% & 129 & 99.72\%\\
\hline
\multirow{15}{*}{LP} & \multirow{5}{*}{BS} & 0 & 105 & 104 & 0.95\% & 104 & 0.95\% & 104 & 0.95\%\\
& & 1 & 945 & 104 & 89\% & 104 & 89\% & 104 & 89\%\\
& & 2 & 8505 & 104 & 98.78\% & 104 & 98.78\% & 104 & 98.78\%\\
& & 3 & 76545 & 104 & 99.86\% & 104 & 99.86\% & 104 & 99.86\%\\
& & 4 & 688905 & 104 & 99.98\% & 104 & 99.98\% & 104 & 99.98\%\\
\cline{2-10}
& \multirow{5}{*}{S} & 0 & 105 & 104 & 0.95\% & 104 & 0.95\% & 104 & 0.95\%\\
& & 1 & 735 & 104 & 85.85\% & 104 & 85.85\% & 104 & 85.85\%\\
& & 2 & 5145 & 104 & 97.98\% & 104 & 97.98\% & 104 & 97.98\%\\
& & 3 & 36015 & 104 & 99.71\% & 104 & 99.71\% & 104 & 99.71\%\\
& & 4 & 252105 & 104 & 99.96\% & 104 & 99.96\% & 104 & 99.96\%\\
\cline{2-10}
& \multirow{5}{*}{K} & 0 & 105 & 104 & 0.95\% & 104 & 0.95\% & 104 & 0.95\%\\
& & 1 & 420 & 104 & 75.24\% & 104 & 75.24\% & 100 & 76.2\%\\
& & 2 & 1680 & 100 & 94.05\% & 100 & 94.05\% & 100 & 94.05\%\\
& & 3 & 6720 & 100 & 98.51\% & 100 & 98.51\% & 100 & 98.51\%\\
& & 4 & 26880 & 100 & 99.63\% & 100 & 99.63\% & 100 & 99.63\%\\
\hline
\multirow{15}{*}{NP} & \multirow{5}{*}{BS} & 0 & 105 & 104 & 0.95\% & 51 & 51.43\% & 13 & 87.62\%\\
& & 1 & 945 & 100 & 89.42\% & 100 & 89.42\% & 100 & 89.42\%\\
& & 2 & 8505 & 100 & 98.82\% & 100 & 98.82\% & 100 & 98.82\%\\
& & 3 & 76545 & 100 & 99.87\% & 100 & 99.87\% & 100 & 99.87\%\\
& & 4 & 688905 & 100 & 99.99\% & 100 & 99.99\% & 100 & 99.99\%\\
\cline{2-10}
& \multirow{5}{*}{S} & 0 & 105 & 104 & 0.95\% & 52 & 50.47\% & 7 & 93.33\%\\
& & 1 & 735 & 104 & 85.85\% & 97 & 86.8\% & 97 & 86.8\%\\
& & 2 & 5145 & 97 & 98.11\% & 97 & 98.11\% & 97 & 98.11\%\\
& & 3 & 36015 & 97 & 99.73\% & 97 & 99.73\% & 97 & 99.73\%\\
& & 4 & 252105 & 97 & 99.96\% & 97 & 99.96\% & 97 & 99.96\%\\
\cline{2-10}
& \multirow{5}{*}{K} & 0 & 105 & 104 & 0.95\% & 52 & 50.47\% & 7 & 93.33\%\\
& & 1 & 420 & 100 & 76.19\% & 100 & 76.19\% & 55 & 86.9\%\\
& & 2 & 1680 & 100 & 94.05\% & 100 & 94.05\% & 100 & 94.05\%\\
& & 3 & 6720 & 100 & 98.52\% & 100 & 98.52\% & 100 & 98.52\%\\
& & 4 & 26880 & 100 & 99.63\% & 100 & 99.63\% & 100 & 99.63\%\\
\hline
\multirow{15}{*}{NA} & \multirow{5}{*} {BS} & 0 & 128 & 127 & 0.78\% & 57 & 55.47\% & 19 & 85.15\%\\
& & 1 & 1152 & 127 & 88.97\% & 127 & 88.97\% & 127 & 88.97\%\\
& & 2 & 10368 & 127 & 98.77\% & 127 & 98.77\% & 127 & 98.77\%\\
& & 3 & 93312 & 127 & 99.86\% & 127 & 99.86\% & 127 & 99.86\%\\
& & 4 & 839808 & 127 & 99.98\% & 127 & 99.98\% & 127 & 99.98\%\\
\cline{2-10}
& \multirow{5}{*}{S} & 0 & 128 & 127 & 0.78\% & 57 & 55.67\% & 19 & 85.15\%\\
& & 1 & 896 & 127 & 85.83\% & 125 & 86.05\% & 90 & 89.95\%\\
& & 2 & 6272 & 127 & 97.97\% & 90 & 98.56\% & 90 & 98.56\%\\
& & 3 & 43904 & 90 & 99.79\% & 90 & 99.79\% & 90 & 99.79\%\\
& & 4 & 307328 & 90 & 99.997\% & 90 & 99.997\% & 90 & 99.997\%\\
\cline{2-10}
& \multirow{5}{*}{K} & 0 & 128 & 127 & 0.78\% & 57 & 55.47\% & 19 & 85.15\%\\
& & 1 & 512 & 127 & 75.2\% & 108 & 78.9\% & 108 & 78.9\%\\
& & 2 & 2048 & 108 & 94.73\% & 108 & 94.73\% & 108 & 94.73\%\\
& & 3 & 8192 & 108 & 98.68\% & 108 & 98.68\% & 108 & 98.68\%\\
& & 4 & 32768 & 108 & 99.67\% & 108 & 99.67\% & 108 & 99.67\%\\
\hline
\multirow{15}{*}{QD} & \multirow{5}{*}{BS} & 0 & 190 & 189 & 0.5\% & 189 & 0.5\% & 189 & 0.5\%\\
& & 1 & 1710 & 189 & 88.95\% & 189 & 88.95\% & 189 & 88.95\%\\
& & 2 & 15390 & 189 & 98.77\% & 189 & 98.77\% & 189 & 98.77\%\\
& & 3 & 138510 & 189 & 99.86\% & 189 & 99.86\% & 189 & 99.86\%\\
& & 4 & 1246590 & 189 & 99.98\% & 189 & 99.98\% & 189 & 99.98\%\\
\cline{2-10}
& \multirow{5}{*}{S} & 0 & 190 & 189 & 0.5\% & 189 & 0.5\% & 189 & 0.5\%\\
& & 1 & 1330 & 189 & 85.8\% & 189 & 85.8\% & 189 & 85.8\%\\
& & 2 & 9310 & 189 & 97.97\% & 189 & 97.97\% & 189 & 97.97\%\\
& & 3 & 65170 & 189 & 99.71\% & 189 & 99.71\% & 189 & 99.71\%\\
& & 4 & 456190 & 189 & 99.96\% & 189 & 99.96\% & 189 & 99.96\%\\
\cline{2-10}
& \multirow{5}{*}{K} & 0 & 190 & 189 & 0.5\% & 189 & 0.5\% & 189 & 0.5\%\\
& & 1 & 760 & 189 & 75.13\% & 189 & 75.13\% & 189 & 75.13\%\\
& & 2 & 3040 & 189 & 93.78\% & 189 & 93.78\% & 189 & 93.78\%\\
& & 3 & 12160 & 189 & 98.45\% & 189 & 98.45\% & 189 & 98.45\%\\
& & 4 & 48640 & 189 & 99.61\% & 189 & 99.61\% & 189 & 99.61\%\\
\hline
\end{tabular}
}
\end{table}

\section{Conclusion}
In this paper, we have introduced a novel technique of error tracing to reduce the required number of error correcting block in any quantum circuit. Pauli errors have been considered. The source of error in any quantum circuit is due to noisy gate and interaction with the environment. We have dealt with these two individually and have provided an efficient algorithm to determine the time for which some qubit(s) are idle while others are undergoing gate operation. The calculation of total error probability for any linear quantum circuit has also been shown.\\

We have, then, considered concatenated codes. The error probability in a linear quantum circuit depends on the gate error and memory error probability. But for concatenated codes, the error probability is largely dependent on the underlying lattice architecture. We have considered the architectures for Bacon-Shor code, Steane Code and Knill Code and have shown the error calculations in each case. The results from the benchmark circuits show that this technique is extremely useful for those technologies where error probability is low, for example, Ion Trap and Superconductor. However, even for other technologies, where this technique does not yield good results in the linear case, a huge amount of savings in the resource is obtainable for concatenated codes. Hence, our results show that it is possible to perform quantum computation in an open environment with much less resource requirement than the ideal case.

\bibliographystyle{plain}       
\bibliography{qip}

\begin{thebibliography}{10}

\bibitem{PhysRevA.73.012340}
Dave Bacon.
\newblock Operator quantum error-correcting subsystems for self-correcting
  quantum memories.
\newblock {\em Phys. Rev. A}, 73:012340, Jan 2006.

\bibitem{bennett1984quantum}
Charles~H Bennett.
\newblock Quantum cryptography: Public key distribution and coin tossing.
\newblock In {\em International Conference on Computer System and Signal
  Processing, IEEE, 1984}, pages 175--179, 1984.

\bibitem{PhysRevLett.70.1895}
Charles~H. Bennett, Gilles Brassard, Claude Cr\'epeau, Richard Jozsa, Asher
  Peres, and William~K. Wootters.
\newblock Teleporting an unknown quantum state via dual classical and
  einstein-podolsky-rosen channels.
\newblock {\em Phys. Rev. Lett.}, 70:1895--1899, Mar 1993.

\bibitem{PhysRevLett.69.2881}
Charles~H. Bennett and Stephen~J. Wiesner.
\newblock Communication via one- and two-particle operators on
  einstein-podolsky-rosen states.
\newblock {\em Phys. Rev. Lett.}, 69:2881--2884, Nov 1992.

\bibitem{doi:10.1137/S0097539796300921}
Ethan Bernstein and Umesh Vazirani.
\newblock Quantum complexity theory.
\newblock {\em SIAM Journal on Computing}, 26(5):1411--1473, 1997.

\bibitem{PhysRevA.54.1098}
A.~R. Calderbank and Peter~W. Shor.
\newblock Good quantum error-correcting codes exist.
\newblock {\em Phys. Rev. A}, 54:1098--1105, Aug 1996.

\bibitem{PhysRevLett.77.3260}
David~P. DiVincenzo and Peter~W. Shor.
\newblock Fault-tolerant error correction with efficient quantum codes.
\newblock {\em Phys. Rev. Lett.}, 77:3260--3263, Oct 1996.

\bibitem{gottesman1997stabilizer}
Daniel Gottesman.
\newblock Stabilizer codes and quantum error correction.
\newblock {\em arXiv preprint quant-ph/9705052}, 1997.

\bibitem{Grover:1996:FQM:237814.237866}
Lov~K. Grover.
\newblock A fast quantum mechanical algorithm for database search.
\newblock In {\em Proceedings of the Twenty-eighth Annual ACM Symposium on
  Theory of Computing}, STOC '96, pages 212--219, New York, NY, USA, 1996. ACM.

\bibitem{KnillNature}
E.~Knill, R.~Laflamme, and G.~J. Milburn.
\newblock A scheme for efficient quantum computation with linear optics.
\newblock {\em Nature}, 409(6816):46--52, Jan 2001.

\bibitem{PhysRevLett.77.198}
Raymond Laflamme, Cesar Miquel, Juan~Pablo Paz, and Wojciech~Hubert Zurek.
\newblock Perfect quantum error correcting code.
\newblock {\em Phys. Rev. Lett.}, 77:198--201, Jul 1996.

\bibitem{6423291}
C.~C. Lin, A.~Chakrabarti, and N.~K. Jha.
\newblock Optimized quantum gate library for various physical machine
  descriptions.
\newblock {\em IEEE Transactions on Very Large Scale Integration (VLSI)
  Systems}, 21(11):2055--2068, Nov 2013.

\bibitem{6565421}
C.~C. Lin, A.~Chakrabarti, and N.~K. Jha.
\newblock Ftqls: Fault-tolerant quantum logic synthesis.
\newblock {\em IEEE Transactions on Very Large Scale Integration (VLSI)
  Systems}, 22(6):1350--1363, June 2014.

\bibitem{1464-4266-7-7-002}
W~J Munro, Kae Nemoto, T~P Spiller, S~D Barrett, Pieter Kok, and R~G
  Beausoleil.
\newblock Efficient optical quantum information processing.
\newblock {\em Journal of Optics B: Quantum and Semiclassical Optics},
  7(7):S135, 2005.

\bibitem{nielsen2010quantum}
Michael~A Nielsen and Isaac~L Chuang.
\newblock {\em Quantum computation and quantum information}.
\newblock Cambridge university press, 2010.

\bibitem{PhysRevA.55.4593}
M.~B. Plenio, V.~Vedral, and P.~L. Knight.
\newblock Conditional generation of error syndromes in fault-tolerant error
  correction.
\newblock {\em Phys. Rev. A}, 55:4593--4596, Jun 1997.

\bibitem{Roos2014}
Christian Roos.
\newblock {\em Quantum Information Processing with Trapped Ions}, pages
  253--291.
\newblock Springer Berlin Heidelberg, Berlin, Heidelberg, 2014.

\bibitem{RevModPhys.82.2313}
M.~Saffman, T.~G. Walker, and K.~M\o{}lmer.
\newblock Quantum information with rydberg atoms.
\newblock {\em Rev. Mod. Phys.}, 82:2313--2363, Aug 2010.

\bibitem{Shor:1996:FQC:874062.875509}
P.~W. Shor.
\newblock Fault-tolerant quantum computation.
\newblock In {\em Proceedings of the 37th Annual Symposium on Foundations of
  Computer Science}, FOCS '96, pages 56--, Washington, DC, USA, 1996. IEEE
  Computer Society.

\bibitem{PhysRevA.52.R2493}
Peter~W. Shor.
\newblock Scheme for reducing decoherence in quantum computer memory.
\newblock {\em Phys. Rev. A}, 52:R2493--R2496, Oct 1995.

\bibitem{spedalieri2008latency}
Federico~M Spedalieri and Vwani~P Roychowdhury.
\newblock Latency in local, two-dimensional, fault-tolerant quantum computing.
\newblock {\em arXiv preprint arXiv:0805.4213}, 2008.

\bibitem{PhysRevLett.77.793}
A.~M. Steane.
\newblock Error correcting codes in quantum theory.
\newblock {\em Phys. Rev. Lett.}, 77:793--797, Jul 1996.

\bibitem{PhysRevLett.91.167005}
Frederick~W. Strauch, Philip~R. Johnson, Alex~J. Dragt, C.~J. Lobb, J.~R.
  Anderson, and F.~C. Wellstood.
\newblock Quantum logic gates for coupled superconducting phase qubits.
\newblock {\em Phys. Rev. Lett.}, 91:167005, Oct 2003.

\bibitem{suchara2013estimating}
Martin Suchara, Arvin Faruque, Ching-Yi Lai, Gerardo Paz, Frederic Chong, and
  John~D Kubiatowicz.
\newblock Estimating the resources for quantum computation with the qure
  toolbox.
\newblock Technical report, DTIC Document, 2013.

\bibitem{Svore:2007:NTF:2011725.2011727}
Krysta~M. Svore, David~P. Divincenzo, and Barbara~M. Terhal.
\newblock Noise threshold for a fault-tolerant two-dimensional lattice
  architecture.
\newblock {\em Quantum Info. Comput.}, 7(4):297--318, May 2007.

\bibitem{PhysRevB.76.035315}
J.~M. Taylor, J.~R. Petta, A.~C. Johnson, A.~Yacoby, C.~M. Marcus, and M.~D.
  Lukin.
\newblock Relaxation, dephasing, and quantum control of electron spins in
  double quantum dots.
\newblock {\em Phys. Rev. B}, 76:035315, Jul 2007.

\bibitem{wilde2013quantum}
Mark~M Wilde.
\newblock {\em Quantum information theory}.
\newblock Cambridge University Press, 2013.

\end{thebibliography}

\end{document}